\documentclass[]{iopart}

\usepackage{iopams}
\usepackage{amsbsy}
\usepackage{latexsym}
\usepackage{graphicx}
\usepackage{url}
\usepackage{acronym}
\usepackage[noabbrev,nameinlink]{cleveref}
\usepackage{url}
\usepackage{aas_macros}

\begin{document}

\title[Localization beyond triangulation]{Localization of transient gravitational wave sources: beyond
triangulation}

\author{Stephen Fairhurst}
\address{Cardiff School of Physics and Astronomy,
Cardiff University, Queens Buildings, The Parade, Cardiff. CF24 3AA
}
\eads{\mailto{fairhursts@cardiff.ac.uk}}

\begin{abstract} 

Rapid, accurate localization of gravitational wave transient events has proved
critical to successful electromagnetic followup.  In previous papers we have
shown that localization estimates can be obtained 
through triangulation based on timing information at the
detector sites.  In practice, detailed parameter estimation routines use 
additional information and provide better localization than is possible based
on timing information alone.  In this paper, we extend the timing based localization
approximation to incorporate consistency of observed signals with two gravitational
wave polarizations, and an astrophysically motivated distribution of sources.  Both
of these provide significant improvements to source localization, allowing many sources
to be restricted to a single sky region, with an area 40\% smaller than predicted by
timing information alone.  Furthermore, we show that the vast majority of sources
will be reconstructed to be circularly polarized or, equivalently, indistinguishable from
face-on.
\end{abstract}

\maketitle

\acrodef{SNR}{signal-to-noise ratio}
\acrodef{GW}{gravitational wave}
\acrodef{EM}{electromagnetic}
\acrodef{PSD}{power spectral density}
\acrodef{BNS}{binary neutron star}

\section{Introduction}
\label{sec:intro}
Numerous gravitational wave signals from binary mergers have now been observed \cite{Abbott:2016blz, 
Abbott:2016nmj, TheLIGOScientific:2016pea, TheLIGOScientific:2017qsa}.  The sky location
of these mergers have been provided with minimal delay, based on low latency \ac{GW}
searches \cite{Messick:2016aqy, Usman:2015kfa, Nitz:2017svb} and rapid sky localization routines 
\cite{Singer:2015ema}.  Subsequently, more detailed parameter
estimation routines have provided updated localizations \cite{Veitch:2014wba}.  The rapid analysis and
 localization enabled the 
\ac{EM} followup of all \ac{GW} events and, most spectacularly, the observation of an \ac{EM} 
counterpart to GW170817-GRB170817A \cite{Monitor:2017mdv} across the 
electromagnetic spectrum \cite{GBM:2017lvd}.

In \cite{Fairhurst2009, Fairhurst:2010is}, it was argued that timing information provides a good
estimate of the localization accuracy of a signal observed in a gravitational wave network.  Nonetheless,
the localization achieved by full parameter estimation analyses \cite{Singer:2015ema, Veitch:2014wba, Singer:2014qca,  
Berry:2014jja} is significantly better than the timing approximation and makes use of the fact that
the observed amplitudes and phases of the signal at different detectors must be correlated \cite{Singer:2015ema}.  
A detailed comparison of the results from 
timing-triangulation and full parameter estimation studies \cite{Grover:2013sha} demonstrated that there 
are important factors above and beyond timing which can improve the localization of sources.
Here, we investigate the physical origin of these improvements and provide an
intuitive explanation of the most important additional factors in localization.
This is achieved by extending the
localization calculation to incorporate correlations between the detectors, due to the fact that there
are only two gravitational wave polarizations, and astrophysical priors on the 
source distributions.  Inclusion of these effects leads to significant improvements in localization.

The fact that a gravitational wave is comprised of two polarizations
guarantees the observed amplitude and phase of the signal at three or more detectors is 
correlated.  Taking this into account improves the localization capabilities of the network.
At sky locations other than the correct location, the requirement of signal consistency
with two \ac{GW} polarizations, reduces the reconstructed network \ac{SNR} which
improves the localization \cite{Wen:2010cr, Wen:2005ui}.
For a three site network, timing triangulation gives two sky locations which are mirror images
in the plane defined by the three detector locations.  Including amplitude and phase consistency can, in principle, 
break this degeneracy and allow us to neglect one of the patches.  In many cases,
the recovered \ac{SNR} at the mirror position will be of lower amplitude than the original,
allowing us to reject the mirror position for around half of the sources.

Gravitational wave sources are expected to be distributed roughly uniformly in volume, at least at the 
sensitive distances of the advanced GW detectors --- large enough to neglect local
inhomogeneities and small enough that rate variations are not dominated by cosmological effects 
or the variation of the star formation rates during the evolution of the universe.  Furthermore, the 
orientation of merging binaries is expected to be uniformly distributed.
Consequently, the majority of sources will be observed close to face-on, as the strongest \ac{GW}
signal is emitted in this direction, and at large distances.  In a three detector network, the signal 
at the mirror sky location will typically be reconstructed at a much smaller distance than the 
true signal, and oriented close to edge on, providing another method of rejecting the mirror sky
location. 

These effects also improve localization within a single sky patch.  At positions offset from the true location, 
the observed \acp{SNR} will not be entirely consistent with a signal, allowing for improved localization.
Additionally, the reconstructed orientation of the source will be
increasingly edge-on and at a smaller distance as we move away from the true position.  
Incorporating amplitude and phase consistency between sites and imposing reasonable
astrophysical priors on the source distribution allows us to localize the vast majority of sources 
to a single patch in the sky.  In addition, the error region for most sources is more accurately
identified by considering the localization of a circularly polarized signal.  For
a binary neutron star system observed in three detectors, the cumulative effect is to provide a factor
 of three improvement in localization over the timing triangulation results.

The majority of the paper deals with the three site advanced LIGO-Virgo network.  
However, we also consider future four and five detector networks incorporating LIGO
India \cite{M1100296, 2013IJMPD..2241010U} and KAGRA \cite{Somiya:2011np}.  
Sources observed in four or more detectors are localized to a single patch in the
sky based on timing alone.  Nonetheless, requiring a consistent signal between the sites and
favoring distant, approximately face-on sources improves localization significantly.

The paper is laid out as follows.  In \sref{sec:beyond_tri}, we briefly recap the timing triangulation
results and then present the extensions to a coherent analysis and the inclusion of astrophysical 
source distribution.  In \sref{sec:results}, we provide the expected localizations for the advanced 
LIGO-Virgo network at design sensitivity as well as four and five detector networks incorporating
KAGRA and LIGO India.  Finally, in \sref{sec:disc} we provide a discussion of the results and indicate
future directions of research.  We provide two appendices: \ref{sec:marg_details} which
describes in detail the process of marginalizing over the astrophysical priors and \ref{sec:loc_details}
which gives a detailed derivation of the localization calculation.

\section{Localization beyond triangulation}
\label{sec:beyond_tri}
In this section, we present the new features which allow for improved localization.  We begin
with a brief review of the timing triangulation \cite{Fairhurst2009, Fairhurst:2010is}, before extending 
the formalism to incorporate correlations between detectors due to a coherent analysis and
the impact of astrophysically motivated priors on the source parameters, particularly location and
orientation of binaries.

\subsection{Timing}
\label{sec:time}

The accuracy with which the time of arrival of a gravitational wave can be measured
in a single detector can be approximated as
\begin{equation}\label{eq:sigma_f}
	\sigma_{t} = \frac{1}{2 \pi \rho \sigma_{f}}
\end{equation}
where $\sigma_{f}$ is the \textit{bandwidth} of the signal in the detector and $\rho$ is
the \ac{SNR} \cite{Fairhurst2009}.  The \ac{SNR} is defined as
\begin{equation}
 	\rho := \frac{(s | h)}{\sqrt{(h | h)}} \, ,
\end{equation}
where $h(t)$ is the template waveform, $s(t)$ is the detector data, 
\begin{equation}\label{eq:inner}
	(a | b) := 4 \, \mathrm{Re} \int \frac{ \tilde{a}(f) \tilde{b}^{\star}(f)}{S(f)}df \, ,
\end{equation}
and $S(f)$ is the \ac{PSD}.  The bandwidth $\sigma_{f}$ is given by 
\begin{equation}
	\sigma_{f}^{2} = \overline{f^2} - \bar{f}^2 
	\quad \mathrm{where} \quad
 	\overline{f^{n}} := \frac{(h | f^{n} h)}{(h | h)} \, .
\end{equation}

Defining $dt_{i}$ as the time difference between the actual time of arrival in a detector and the recovered 
arrival time, we can obtain the probability distribution for $\mathbf{dt} = (dt_{1}, \ldots, dt_{N})$ which is 
proportional to the likelihood evaluated at those time offsets.  Considering timing alone, the time delay 
distributions in each detector are independent and we obtain the distribution for the time-delay vector 
$\mathbf{dt}$ as,
\begin{equation}\label{eq:timing_prob}
	p(\mathbf{dt}) \propto \exp\left\{- \frac{1}{2} \left[\sum_{i} 
	4\pi^{2} \sigma_{f,i}^{2} \rho_{i}^{2} dt_{i}^{2} ) \right] \right\} \, .
\end{equation}
This is converted to localization accuracy by re-expressing the relative time delays in terms of sky locations, 
as discussed in detail in \cite{Fairhurst2009, Fairhurst:2010is} and in \ref{sec:loc_details}.
We are ignoring the correlation of the
arrival time with other parameters, such as the component masses of a binary merger.  
Since these correlations are similar in all detectors, they will have minimal effect on the 
measured time delay between sites, which is the important quantity for localization 
\cite{Fairhurst2009, Singer:2014qca}.

\subsection{Coherent Network Analysis}
\label{sec:coherent}

Since a gravitational wave signal is comprised of only two polarizations, with a network of
three or more detectors there are consistency requirements between the observed \acp{SNR}
in the detectors that comprise the network.  As discussed in \cite{Harry:2010fr}, we can express the 
network \ac{SNR} as the projection of the individual detector \acp{SNR} onto the two-dimensional
signal space, namely the two dimensional physical subspace of signals which is consistent
with the two polarizations of gravitational waves.  

To do so, we must first introduce the \textit{complex} \ac{SNR} Z.  
This is the output of the matched filter of a template waveform $h$ placed at some fiducial distance
$D_{0}$, with two orthogonal phases 
$h_{0}$ and $h_{\frac{\pi}{2}}$, against the data $s$,
\begin{equation}\label{eq:complex_snr}
	Z = \frac{ (s | h_{0}) + i (s| h_{\frac{\pi}{2}}) }{\sigma_{h}} 
	\quad \mathrm{where} \quad
	\sigma_{h}^{2} = (h_{0} | h_{0}) \, .
\end{equation}
Then $\rho = |Z|$ gives the \ac{SNR} maximized over the signal amplitude and phase.
The network \ac{SNR} can be expressed as 
\begin{equation}\label{eq:coh_snr}
\rho_{\mathrm{net}}^{2} = Z_{i}^{\star} P^{ij} Z_{j} \, .
\end{equation}
where $P^{ij}$ is a rank 2 projection matrix that projects the \ac{SNR} onto the two
gravitational wave polarizations and there is an implicit sum over $i$ and $j$ in \eref{eq:coh_snr}.  
When the observed \acp{SNR} are consistent
with a signal, $P^{ij} Z_{j} = Z^{i}$.  It is often convenient to work in the dominant polarization frame,
where the network is maximally sensitive to the $+$ polarization,\footnote{Specifically, we choose a polarization angle that maximizes $| w_{+} |$.  It can be shown that as
a consequence $w_{+} \cdot w_{\times} = 0$ and $ | w_{\times} | $ is minimized. 
See e.g.~\cite{Harry:2010fr, Klimenko:2011} for details.}
and the projection operator is then given by \cite{Harry:2010fr}
\begin{equation}\label{eq:projection}
	P^{ij} =  \left(
	\frac{ w^{i}_{+} w^{j}_{+} }{ | w_{+}|^{2} } +  \frac{w^{i}_{\times} w^{j}_{\times} }{ | w_{\times}|^{2} }
	\right) 
	\quad \mathrm{where} \quad 
	w^{i}_{+,\times} = \sigma_{h}^{i} F^{i}_{+, \times}
	\, .
\end{equation}
The quantities $w^{i}_{+,\times}$ are the sensitivity weighted detector response functions, and $F^{i}_{+,\times}$ are
the well known detector response functions, see e.g. \cite{thorne.k:1987} for the definition.

Consider the situation where the template matches the signal in the detector data, apart from a time and phase offset,
and neglect the effect of noise contributions.  To calculate the \ac{SNR}, we simply note that a time offset can 
be expressed in the frequency domain as $\exp[2\pi i f dt]$ and expand to second order in $dt$.   Following 
\cite{Fairhurst2009}, we have
\begin{equation}\label{eq:z_dt}
Z_{i}(dt_{i}) = Z_{i} \left[ 1 + 2\pi i \bar{f}_{i} dt -  2 \pi^{2} \overline{f^{2}_{i}} dt_{i}^2 \right] \, .
\end{equation}

Substituting this expansion into \eref{eq:coh_snr}, we obtain an expression for the network \ac{SNR} for
a given set of time delays $\mathbf{dt}$.  The likelihood, maximized over the amplitude and phase
of the signal in the two polarizations, is given by
\begin{equation}
	\Lambda_{\mathrm{max}} = \exp\left[ \frac{\rho_{\mathrm{net}}^{2}}{2}\right] 
\end{equation}
as discussed in detail in \ref{sec:marg_details}.  The associated posterior probability distribution 
for $\mathbf{dt}$ is proportional to the likelihood so that
\begin{eqnarray}\label{eq:net_prob}
	p(\mathbf{dt}) &\propto& \exp\left\{ - \frac{1}{2} 
	\left[ \sum_{i} 4\pi^{2} \sigma_{f,i}^{2}  \rho_{i}^{2} dt_{i}^{2} \right. \right. \\
  && \qquad \left. \left.
  +  \sum_{i,j} 4\pi^{2} \bar{f_{i}} dt_{i} Z^{\star}_{i} (\delta^{ij} - P^{ij}) \bar{f_{j}} dt_{j} Z_{j} \right] \right\} \nonumber
\end{eqnarray}
The first term is identical to the timing triangulation result
\eref{eq:timing_prob}, while the second term imposes amplitude and phase consistency
between the sites.  Indeed, the timing expression is recovered, as expected, by setting 
$P^{ij} = \delta^{ij}$.  We can intuitively understand the origin of the additional term as follows:
when localizing using only time delays, we are free to maximize individually the amplitude
and phase of the waveform in each of the detectors; in the coherent network analysis, we are
restricted to waveforms consistent with two \ac{GW} polarizations --- any violation of this will cause
a loss in SNR as power is removed from the signal space $P^{ij}$ and deposited in the
null space, $N^{ij} = \delta^{ij} - P^{ij}$, which, physically, is not expected to contain any GW 
power \cite{GuerselTinto1989, Wen:2005ui}.   

The result \eref{eq:net_prob} has been obtained under two significant
approximations. First, we have kept terms only up to quadratic order in $dt$.  This quadratic approximation
is accurate for moderate or large \acp{SNR} as discussed in \cite{Fairhurst2009}.  Second, we have
neglected the fact that the detector response $F_{+, \times}$ varies across the sky and just used the
value at the actual signal location.  This can be justified by considering a point offset from the true location by
an angle $\delta \phi$.
Since the antenna patterns are quadrupolar, the response will characteristically change at a rate 
$2 \delta \phi$.  Meanwhile,
the time offset at a detector will be $\delta t \sim (\Delta t) \delta \phi$ where $\Delta t$ is the 
separation between detectors (on the order of 10s of ms).  Therefore, the fractional 
fractional change in SNR will be $(2\pi \bar{f} \Delta t) \delta \phi \sim 10 \delta \phi$, for a typical value 
of $\bar{f} = 100$ Hz.  The effect on phasing is several times larger than the effect
of the antenna response, so we ignore the latter it in the calculation that follows. 
For points on the sky where signals are 
poorly localized, in particular those close to the plane defined by the detectors, the changing antenna 
response can be important, although the localization will still be poor. The importance
of the changing antenna response has been considered in detail in, e.g., \cite{Keppel:2013uma, Wen:2010cr}.

To get a sense of the expected improvement from coherent localization, it is useful to look at 
a simple example.  Consider the situation where we have three detectors which are equally sensitive
to a signal and that two detectors are sensitive to the $+$ polarization while the third is
sensitive to $\times$.  To a reasonable approximation, the two LIGO detectors are aligned, and
consequently sensitive to one \ac{GW} polarization while, for many sky locations, Virgo is
sensitive to the other \ac{GW} polarization.  Thus, we assume $w^{H}_{+} = w^{L}_{+} = w^{V}_{\times}$
(and $w^{H}_{\times} = w^{L}_{\times} = w^{V}_{+} = 0$).  In this case the projection matrix $P_{ij}$ 
(where the row/column order is H, L, V) is given by
\begin{equation}
P_{ij} = \left(  \begin{array}{cccc}
\frac{1}{2} & \frac{1}{2} & 0 \\
\frac{1}{2} & \frac{1}{2} & 0 \\
0 & 0 & 1\end{array}\right) \, .
\end{equation}

We are interested in the localization of a source, so we want to see how $p(\mathbf{dt})$ falls off as we
move away from the true location.  The geometry of the LIGO-Virgo 
network can be approximated by an isosceles triangle, with the LIGO detectors separated by $10$ms and the 
light travel time between each LIGO site and Virgo around $26$ms.  
First, consider recovering the source with time offsets 
$dt_{H} = dt_{L} = -dt_{V}$, i.e. at a position offset from the true location along a line connecting the 
midpoint between the LIGO detectors to Virgo.  Since $dt_{H} = dt_{L}$ and $Z_{H} = Z_{L}$
the second line of \eref{eq:net_prob} vanishes and the localization matches the timing result.  
If, instead, we have $dt_{H} = - dt_{L}$ with $dt_{V} = 0$, i.e. offset from the true position along a line 
connecting the two LIGO detectors, then the probability distribution is
\begin{eqnarray}
	p(\mathbf{dt}) &\propto& \exp\left\{ - \frac{1}{2} 
	\left[ \sum_{i=H, L} 4\pi^{2} (\sigma_{f,i}^{2} + \bar{f_{i}}^{2}) \rho_{i}^{2} dt_{i}^{2} \right] \right\} \, .
\end{eqnarray}
This leads to a characteristic width of $(2\pi\rho f_{\mathrm{rms}})^{-1}$, where 
$f_\mathrm{rms} = (\sigma_{f,i}^{2} + \bar{f_{i}}^{2})^{1/2}$,  rather than $(2\pi\rho \sigma_{f})^{-1}$.
The improvement arises from the fact that the two LIGO detectors are required to observe
the same polarization of \ac{GW} and hence the same amplitude and phase.  For the advanced LIGO 
detectors at design, 
$\sigma_{f} \approx 120 \mathrm{Hz}$, $\bar{f}  \approx 100 \mathrm{Hz}$ and
$f_{\mathrm{rms}} \approx 155 \mathrm{Hz}$ for \ac{BNS} signals.  For this example, 
the localization area will be reduced by around $25\%$.

\subsection{Mirror Sky Locations}
\label{sec:mirror}

For a gravitational wave detector network comprising three detectors at different sites, 
localization based on timing will lead to two possible locations, one above and the other below
the plane defined by the locations of the detectors.  From timing information alone, it is not possible to 
distinguish  
the correct location from the \textit{mirror} location.  However, once we move to coherent localization, 
it is, in principle, possible to distinguish between these positions.  
In the mirror location, the observed amplitudes and phases of the signal will generally not be 
consistent with a gravitational wave signal comprising only two polarizations --- a fraction of the 
power will be in the null stream \cite{GuerselTinto1989}, thereby reducing the network \ac{SNR} and allowing
us to reject the mirror position.

At the mirror sky location, the measured network \ac{SNR} will be
\begin{equation}
  \rho_{\mathrm{mirror}}^{2} = Z^{\star}_{i} {P}^{ij}_{\mathrm{mirror}} Z_{j}
\end{equation}
where $Z_{i}$ is the complex \ac{SNR} in detector $i$ and ${P}^{ij}_{\mathrm{mirror}}$ is the projection onto the
signal space appropriate for for the mirror position.  Typically, the observed \acp{SNR} will not be
consistent with the mirror location so ${\rho}_{\mathrm{mirror}} < \rho_{\mathrm{net}}$.\footnote{When the observed \acp{SNR} in the detectors at the mirror location are not consistent with a gravitational wave, 
a nearby sky location may give a larger network \ac{SNR} even though the \acp{SNR} in the
single detectors will be lower.  This is discussed in detail in \ref{sec:loc_details}.  
In the following discussion, we use $\rho_\mathrm{mirror}$ to denote
the maximum \ac{SNR} near the mirror location.}

There are two peaks in the sky location distribution --- one at the true location and the other at the mirror location. 
However, since the SNR associated to the mirror location is smaller than at the correct location, the
peak will be lower and this may allow us to reject it.  
To identify a confidence region containing a fraction $p$ of
the posterior, we typically find the minimum area that contains the required
probability.  Equivalently,  start with the peak of the distribution and add points with the
highest available probability until we have incorporated the appropriate fraction of the probability.
Thus, if the second peak is low enough, it will not be included in the confidence region.

In many cases, the width of the localization distribution, obtained from \eref{eq:net_prob}, at the true and 
mirror locations is comparable.  If we assume that to be
the case, then the probability distribution is well approximated by two 2-dimensional Gaussian
distributions of equal widths, and the relative height of the second peak is given by 
\begin{equation}
	\exp\left[ -\frac{\Delta \rho^{2}}{2}\right] \quad \mathrm{where} 
	\quad  
	\Delta \rho^{2} = \rho_{\mathrm{net}}^{2} - \rho_{\mathrm{mirror}}^{2} \, .
\end{equation}
Then, the second peak will not appear in the localization area provided a fraction $p$ of the total probability
lies in the first region, with a likelihood greater than the second peak, i.e
\begin{equation}
	1 - \exp\left[-\frac{\Delta \rho^2}{2}\right] > p \left(1 + \exp\left[- \frac{\Delta \rho^{2}}{2}\right] \right)
\end{equation}
Here, the left hand side is the integrated probability in the first peak down to the level of the 
second peak; the right hand side is a $p$ times the total integrated probability.  
Thus, if 
\begin{equation}
\Delta \rho^2 > 2 [\ln(1+p) - \ln(1-p)]
\end{equation}
the source will be localized to a single sky patch.  

To obtain a single sky patch at 90\% confidence requires $\Delta \rho^2 > 5.9$.  For a signal
at \ac{SNR} of 12, this equates to a difference in SNR of 0.25 or greater between the true 
and mirror position.  At 50\% confidence, this reduces to $\Delta \rho^2 > 2.2$ or a difference 
in SNR of less than 0.1.  For the advanced LIGO-Virgo detector network at design, about 60\% of 
sources are localized to a single patch at 90\% confidence based on this criterion. 

\subsection{Distribution of Sources}
\label{sec:priors}

Gravitational wave events from binary mergers are reasonably assumed to be uniformly 
distributed in volume, at least at distances on the scale of hundreds of mega parsecs.\footnote{For smaller distances, the discrete structure of galaxies in the nearby universe will
affect the distribution \cite{LIGOS3S4Galaxies} while at larger distances, cosmological evolution
and the variation of the merger rate with redshift have to be taken into account \cite{TheLIGOScientific:2016htt}.} Binary mergers are also expected to be uniformly distributed in source orientation.  So far, we have not taken 
these factors into account when calculating the localization areas.  The expression in equation \eref{eq:coh_snr}
is the maximized \ac{SNR} --- with maximization performed over the amplitude and phase of the two
\ac{GW} polarizations or, equivalently, the distance to and orientation of the binary.  
In recovering estimates of parameters, it is typical to marginalize (rather than maximize) over any unwanted, or
nuisance, parameters.  This can have a significant impact upon the localization results.

\begin{figure*}
\centering
\includegraphics[height=.45\textwidth]{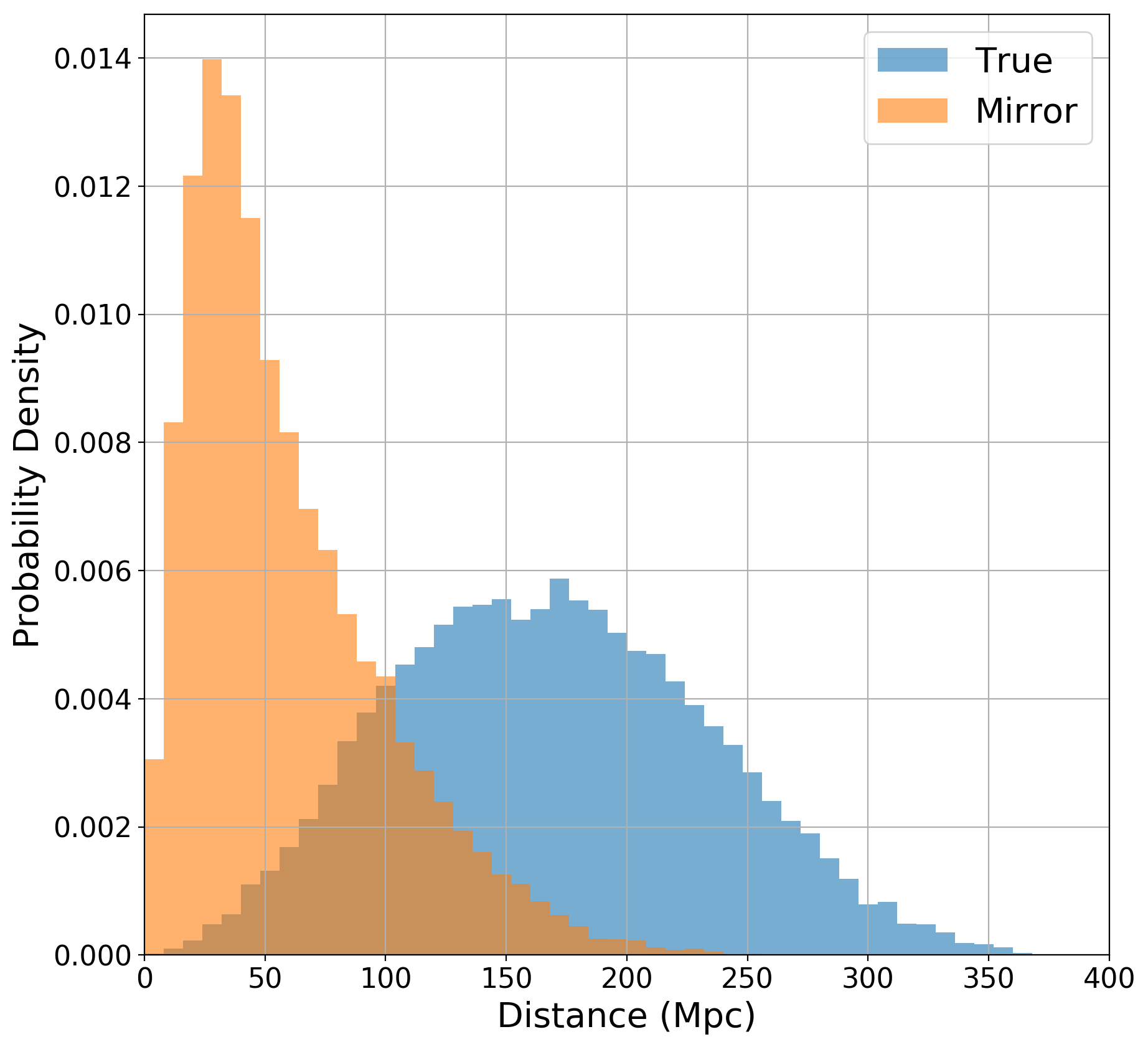}
\includegraphics[height=.45\textwidth]{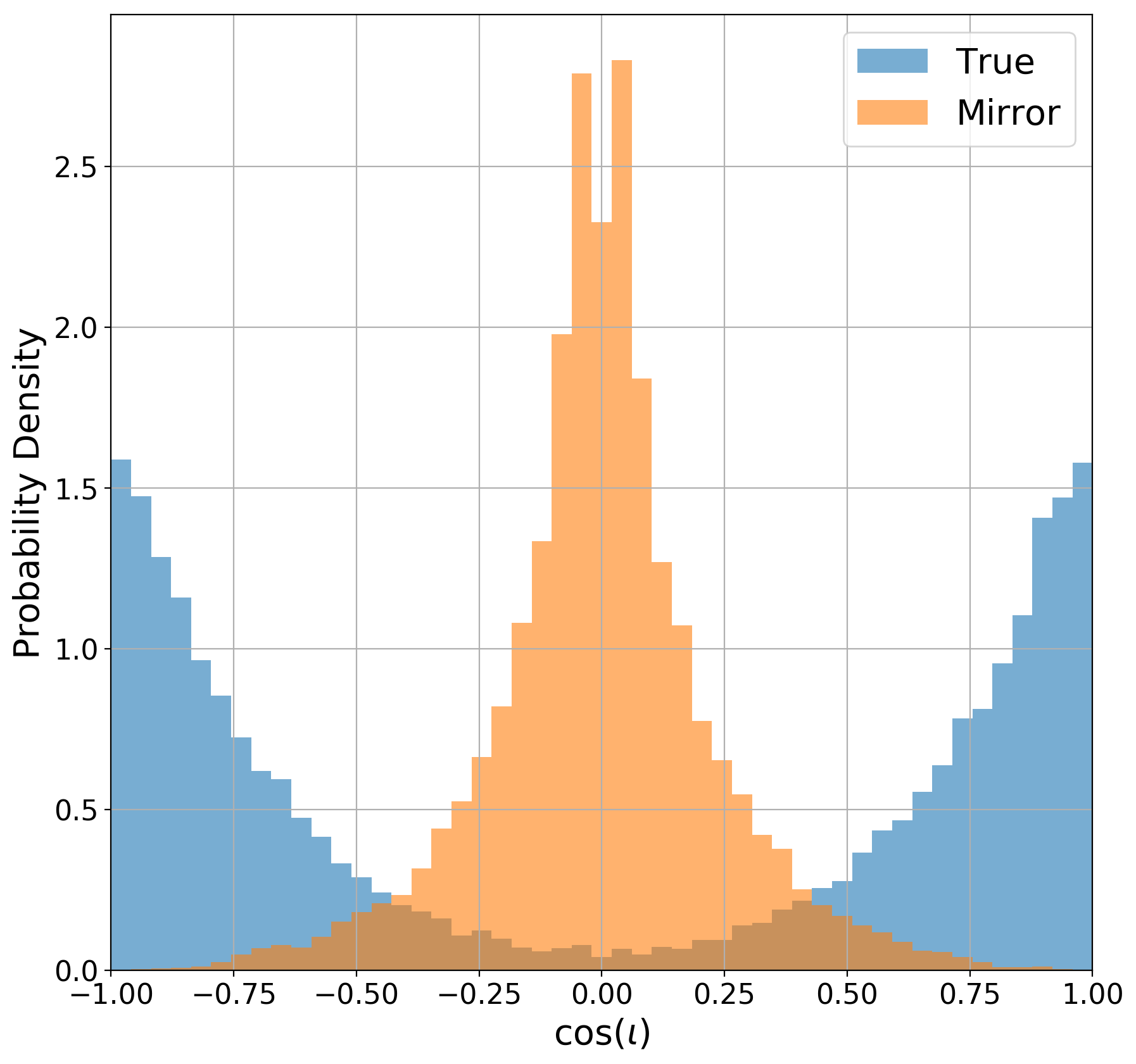}
\caption{Distribution of inferred distance and inclination, $\iota$, of observed sources at the true sky location 
and the mirror position.  The majority of sources at the true location are observed to be face-on and
at distances between 100 and 200 Mpc.  The majority of sources reconstructed at the mirror location
are close to edge on and at distances between 10 and 100 Mpc.}
\label{fig:mirror_di}
\end{figure*}

First, consider our ability to exclude the mirror location.
For a given sky location, the observed \acp{SNR} and detector sensitivities encoded 
in $w_{+, \times}$ can be used to infer the distance to, and orientation of, the source.  The distribution of inferred 
distances for a population of \ac{BNS} signals that might be observed by the Advanced LIGO-Virgo
network operating at design sensitivity is shown in  \fref{fig:mirror_di}. This is obtained by generating sources 
uniform in 
volume and binary orientation and declaring an event to be observed if the expected network \ac{SNR} 
is greater than 12 and the \ac{SNR} in at least two detectors is greater than 5.
The distributions for the correct and mirror locations differ
significantly, with the true distance distribution peaking around $150$ Mpc and the majority
of sources close to face-on.  For the mirror location, the inferred distances are, on average, a factor of
three smaller, and the majority of sources are recovered edge on.  

Initially, let us only consider the distance.  The fractional uncertainty in the distance measurement is 
inversely proportional to the observed SNR, i.e.
\begin{equation}
\frac{\Delta D}{D} \propto \frac{1}{\rho_{\mathrm{net}}} \, .
\end{equation}
Assuming that signals are uniformly distributed in volume, the likelihood of observing a signal
at a distance $D$ is proportional to $D^{2} \Delta{D} \propto D^{3}/\rho_{\mathrm{net}}$.  
Thus, a simple marginalization over the distance leads to a re-weighting of the likelihood as
\begin{equation}
\Lambda_{D} \propto \left(\frac{D^{3}}{\rho_{\mathrm{net}}} \right) \Lambda_{\mathrm{max}} \, . 
\end{equation}
Then, following a similar argument to the one presented in \sref{sec:mirror}, it follows that if the inferred distance at the 
mirror sky location is a factor of $8/3$ lower than the true location, then it will be excluded from the 90\% localization, even if the maximized likelihood is identical.\footnote{In addition, signals are statistically more likely to come from regions with greater sensitivity and,
consequently, on average, the mirror location will be less sensitive.  Thus, simply weighting
the two positions based on the detectors' sensitivity can increase the fraction of sources
localized to a single region.  However, this is only a minor effect.}
A simple distance-based scaling of the likelihood, combined with the reduced \ac{SNR} at the mirror location
discussed in \sref{sec:coherent},
increases the fraction of sources localized to a single patch with 90\% confidence from 60\% to 75\%.

\subsection{Face-on sources}
\label{sec:face_on}

Assuming reasonable parameter priors can also improve the localization accuracy for a
single sky location. To illustrate this, consider again a network with three detectors equally
sensitive to a given source, with two (H and L) sensitive to the $+$ polarization and the third (V)
to $\times$, so that  $w^{H}_{+} = w^{L}_{+} = w^{V}_{\times}$.  In addition, assume that
the source is face-on, so that the observed gravitational wave is circularly polarized and 
$Z_{H} = Z_{L} = - i Z_{V}$.  If the source is reconstructed at a sky location
corresponding to time offsets $dt_{H} = dt_{L} = - dt_{V}$ then, it follows from \eref{eq:z_dt} 
that, the recovered phase in the H and L detectors will be $2 \pi \bar{f} dt$ while
in $V$ it will be $(\frac{\pi}{2} - 2 \pi \bar{f} dt)$.  

\begin{figure*}
\centering
\includegraphics[width=.7\textwidth]{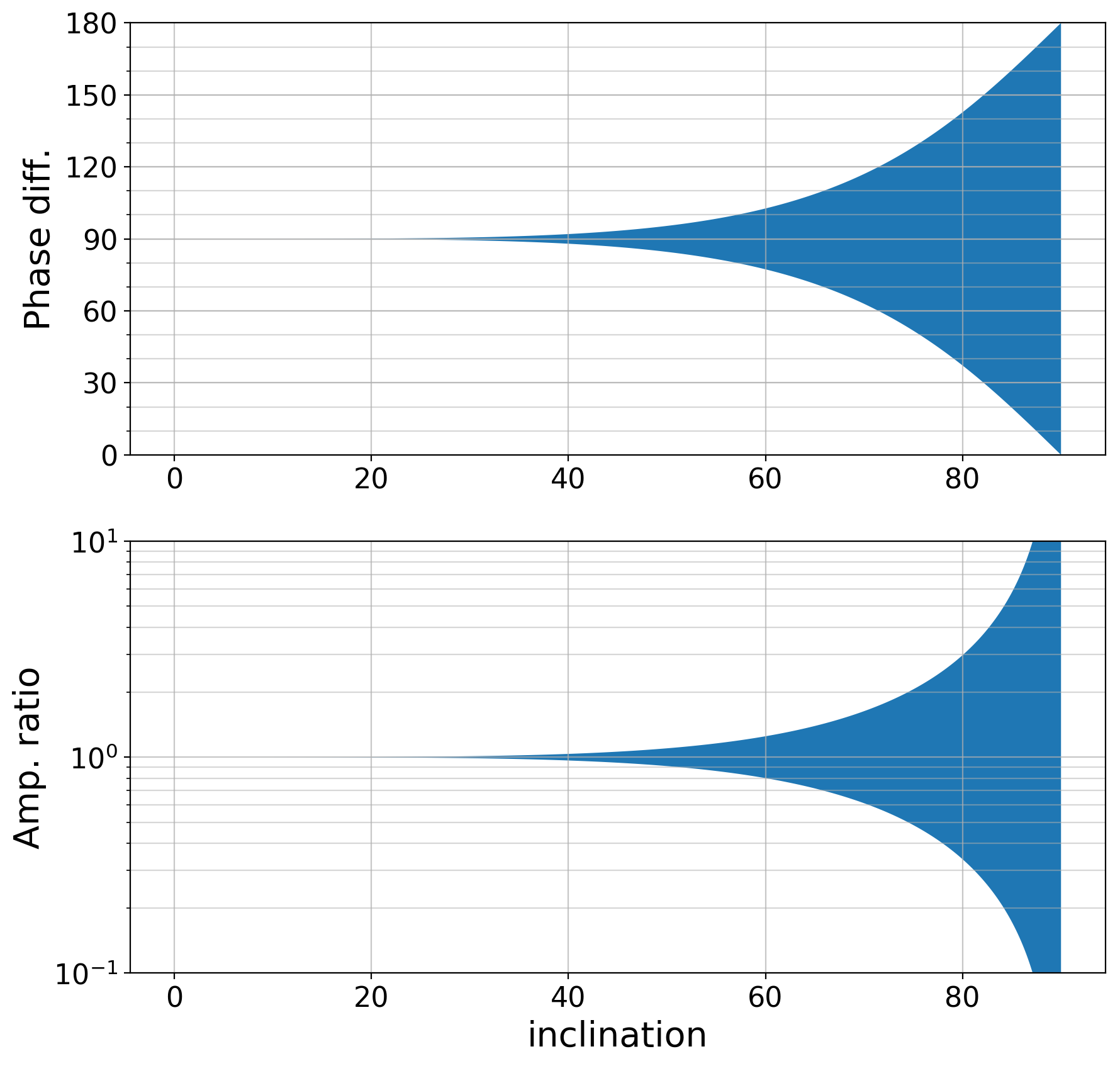}
\caption{The allowed phase difference between and amplitude ratio of the $+$ and $\times$ gravitational wave 
polarizations as a function of the binary inclination.  For approximately face-on signals, the phase difference
is restricted to be very close to $90^{\circ}$ while the amplitude ratio is essentially unity.  A $10^{\circ}$
difference in phase or a $10\%$ difference in amplitude, requires an inclination angle greater than $50^{\circ}$.
At an inclination of $90^{\circ}$ any amplitude ratio and phase difference is possible.}
\label{fig:inclination}
\end{figure*}

In \fref{fig:inclination} we show the range of possible phase and amplitude differences between the
observed $+$ and $\times$ polarizations as a function of the binary inclination.  For edge on binaries, the
emitted \ac{GW} is linearly polarized so, depending upon the polarization angle, any amplitude ratio and
phase difference is possible between $+$ and $\times$.  For face-on signals, the waveform is circularly
polarized so the phase difference is required to be $\pm 90^{\circ}$, with equal signal amplitude
in the two polarizations.  For sources close to face-on, the amplitudes of the two polarizations must
be similar and the phase difference close to $90^{\circ}$.  Interestingly,
the inclination must be large before any significant difference from the face-on case is visible.  In
particular, a $10^\circ$ phase difference or $10\%$ amplitude difference is only possible if the binary 
inclination is greater than $50^\circ$.

\begin{figure*}
\centering
\includegraphics[width=.7\textwidth]{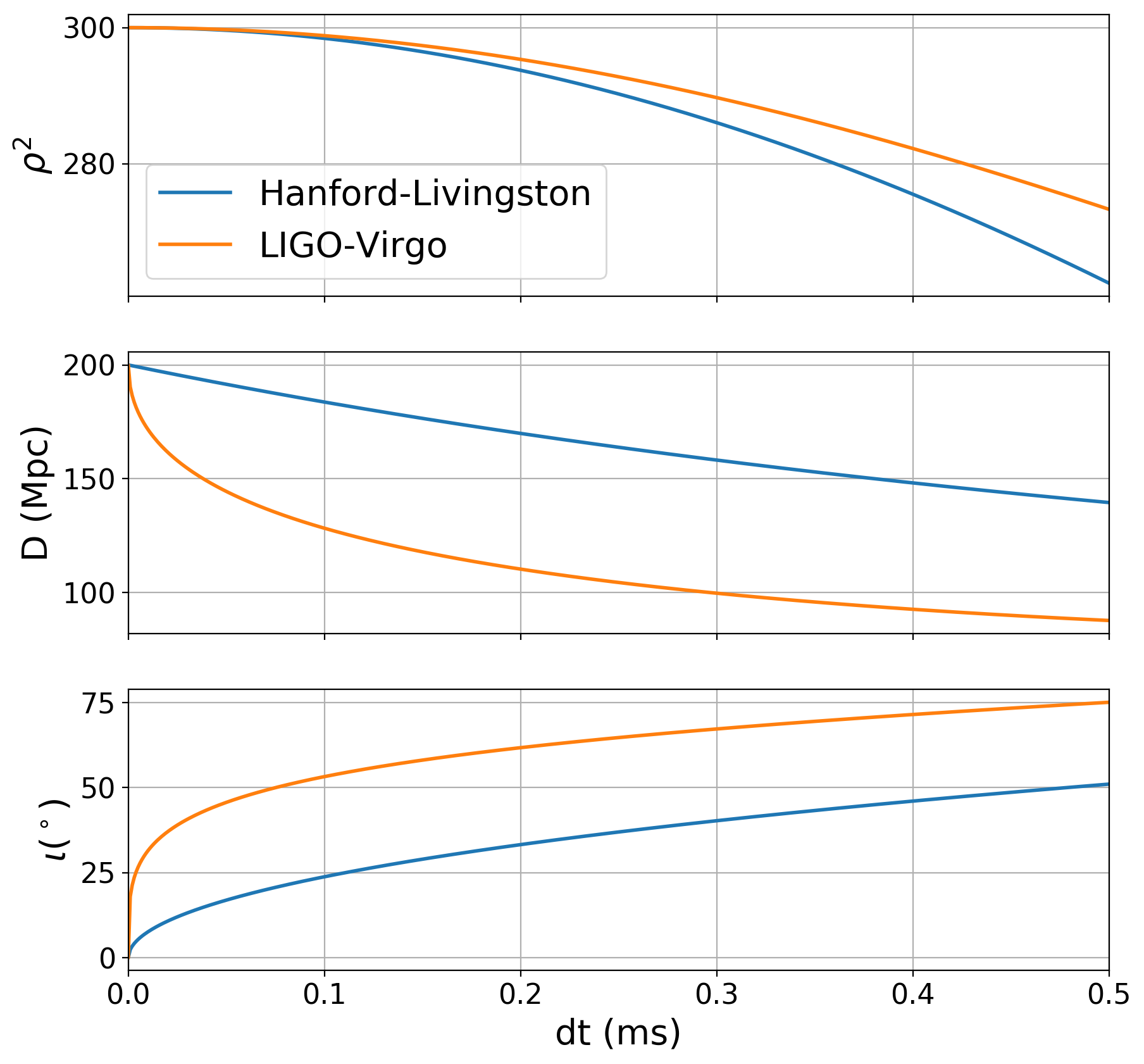}
\caption{The top panel shows the falloff in \ac{SNR} as a function of timing offset.  
We assume a face-on source which gives \ac{SNR} 10 in all detectors with the two LIGO detectors 
sensitive to the $+$ polarization, and Virgo to the $\times$ polarization.  The middle panel
shows the reconstructed distance and bottom panel the inclination angle $\iota$ as a function of the 
timing offset.  For time offsets between LIGO-Virgo, the \ac{SNR} falls off more slowly but, for even 
a small timing error, the maximum \ac{SNR} corresponds to a binary at a smaller distance
which is significantly inclined.}
\label{fig:time_offset}
\end{figure*}

In \fref{fig:time_offset}, we show the maximized \ac{SNR} as a function of the timing offset between the 
true and reconstructed location, as well as the inferred distance and inclination.  
This is plotted for both an offset between the two LIGO sites,
and an offset between LIGO and Virgo.  As expected from \sref{sec:coherent}, the \ac{SNR} falls of more 
rapidly with a timing error between the two LIGO sites, as they are sensitive to the same polarization.  
For changes in the time delay between the LIGO sites, the source
is recovered at a slightly reduced distance and not exactly face-on.  The inferred
amplitude of \ac{GW} in the $+$ polarization will be slightly reduced due to the phase offset between
the signal in the two LIGO detectors.  For changes in the LIGO-Virgo time delay, the maximum \ac{SNR} 
corresponds to a signal which is significantly closer than the true source and highly inclined. 
Already with a $0.05$ ms offset the reconstructed distance is reduced by $25\%$ and the reconstructed 
inclination is $40^{\circ}$.  As discussed in \sref{sec:priors}, a
lower inferred distance reduces the likelihood and this serves to improve localization.

\subsection{Astrophysical Priors}
\label{sec:marg_like}

To provide a more accurate assessment of the impact of distance and orientation priors on the source
reconstruction, we need to calculate the likelihood marginalized over these degrees of freedom.  As discussed
in \sref{sec:priors}, this will give higher weight to sources at a greater distances.  It will also give higher weight to sources
which are close to face-on or face-off, as the observed waveform varies only slowly with
inclination, and the polarization and phase of the signal are degenerate for face-on sources.

In \ref{sec:marg_details}, we derive an expression for the approximate marginalized likelihood of a signal 
based on the observed complex \acp{SNR} and detector sensitivities encoded in $w_{+, \times}$.
The marginalized likelihood can be approximated by two contributions.  The first is simply the 
area of the peak around the maximum 
likelihood \cite{Prix:2009tq, Whelan:2013xka}.  However, when the binary is close to 
circularly polarized, this does \textit{not} give a good approximation to the marginalized result as there 
is a significant volume of parameter space associated with approximately
face-on binaries, for which the waveform is circularly polarized, that provides a large contribution to the 
likelihood.  Thus,
we incorporate a second contribution based on the likelihood for approximately circularly polarized 
waveforms.  The total marginalized likelihood is the sum of these two contributions.  

To illustrate the result, we return to the example discussed previously, namely a network of three 
detectors equally sensitive to the signal, all observing with \ac{SNR} 10.  We vary the
inclination angle of the source, while keeping the network \ac{SNR} fixed, and calculate the
marginalized likelihood.  To make a connection with the previous results, we re-express the 
marginalized likelihood in terms of a re-weighted \ac{SNR} as 
\cite{Nitz:2017svb}
\begin{equation}
	\tilde{\rho}^{2}(\iota) = \rho^{2} + 2 \log \left( \frac{\Lambda_{\mathrm{marg}}(\iota)}{
	\hat{\Lambda}_{\mathrm{marg}}} \right) 
\end{equation}
where $\Lambda_{\mathrm{marg}}(\iota)$ is the marginalized likelihood for a source with
inclination $\iota$, and
$\hat{\Lambda}_{\mathrm{marg}}$ is the greatest value of the marginalized likelihood.
This allows us to
equate an unfavorable orientation with and effective reduction of \ac{SNR}.

\begin{figure*}
\centering
\includegraphics[width=.7\textwidth]{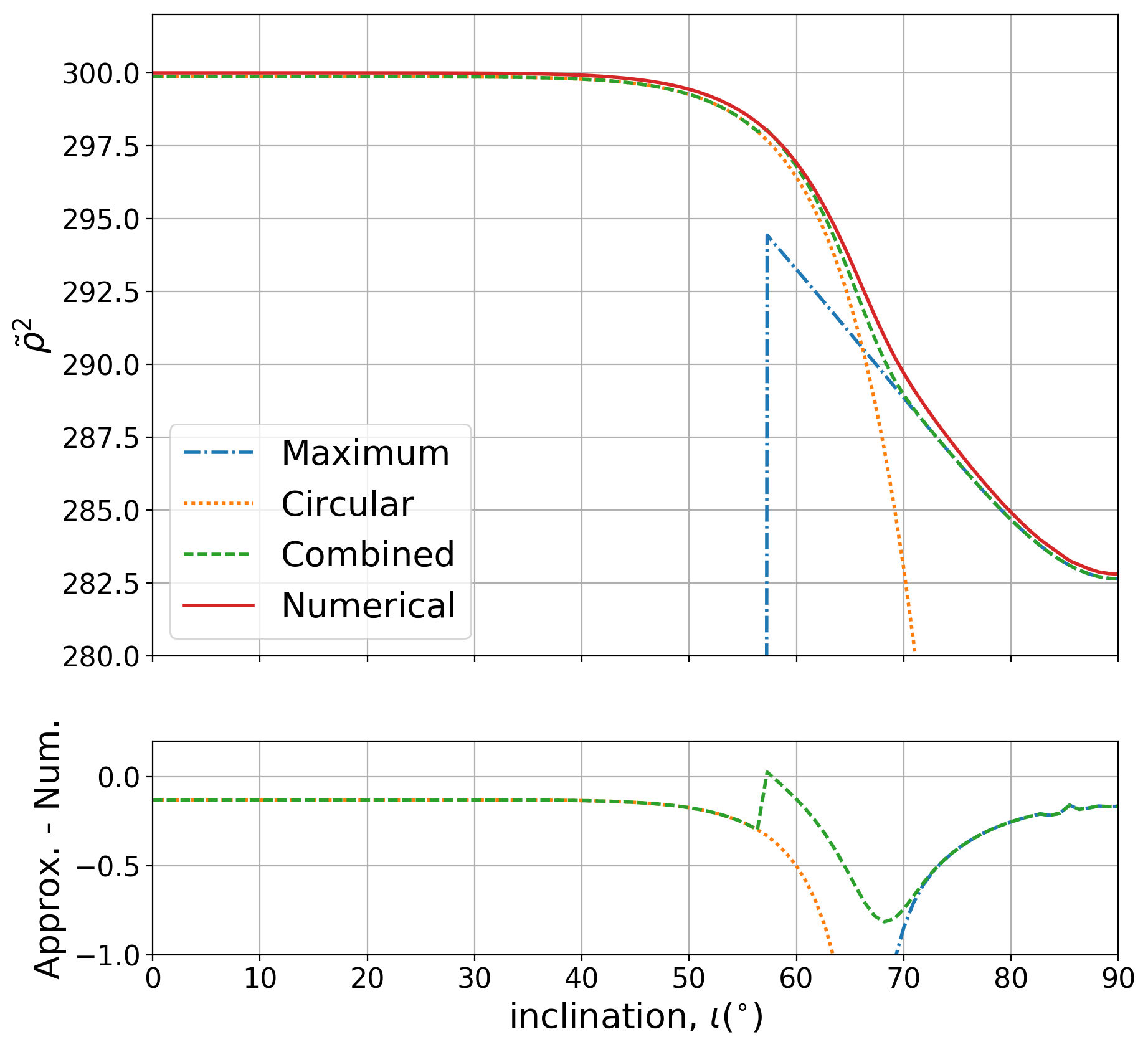}
\caption{The top panel shows $\tilde{\rho}^{2}$, the square of the network \ac{SNR} 
re-weighted by marginalizing over the distance 
and orientation parameters, as a function of the binary inclination.  The maximum is obtained for signals which are
face-on, while for edge on systems $\tilde{\rho}^{2}$ is reduced by over 5\%.  The re-weighted \ac{SNR}
is calculated by numerical integration and also by calculating the approximate contributions from around the
maximum likelihood and by restricting to circularly polarized signals.  For signals with inclination
less than $60^{\circ}$ it is only the circular polarization contribution that matters, 
for inclination above $70^{\circ}$ only
the maximum likelihood is important and between $60^{\circ}$ and $70^{\circ}$ both contribute.  The 
bottom panel shows the difference between the approximate and the numerical result. The approximate
result gives an error in $\tilde{\rho}^{2}$ less than unity for all values of inclination.}
\label{fig:marg_like}
\end{figure*}

In \fref{fig:marg_like} we show the re-weighted \ac{SNR} computed
both numerically and with the approximation based on maximum likelihood and circular polarized signals 
described above.  For face-on, or close to face-on signals,
the re-weighted \ac{SNR} is unchanged.  These are the signals which are consistent with the largest
volume of parameter space.  For edge on systems, $\tilde{\rho}^{2}$ reduces by almost 20,
which is equivalent to reducing the \ac{SNR} in each detector from 10 to 9.7.  Equivalently,
the chance of observing a face-on system is over a thousand times higher than an edge on system.

For each binary, we can evaluate whether the likelihood is predominantly
accumulated around the maximum \ac{SNR} or is associated with circularly polarized emission.  If it is
the former, then the coherent localization presented in \sref{sec:coherent} is appropriate.  However,
if the likelihood is dominated by circularly polarized signals, the localization can be improved
by imposing this restriction.  To do so, we use the previous
localization results but now project onto the space of circularly polarized signals using 

\begin{equation}\label{eq:circ_project}
P^{ij}_\mathrm{circ} =  \left[
\frac{(w^{i}_{+} \pm i w^{i}_{\times})^{\star} (w^{j}_{+} \pm i w^{j}_{\times})}{
  |w_{+}|^2 + |w_{\times}|^2 }  \right] \, ,
\end{equation}
where the $+/-$ signs give the left and right circular polarizations, respectively.
This, in effect, restricts the signal to a single, overall free phase, as conjectured in 
\cite{Grover:2013sha}.  In the space of time-delays, there is now a single direction ---
the one that leaves the relative phase between all pairs of detectors unchanged --- where
the localization is the same as the timing result.  For all other directions, the localization
width is $(2\pi\rho f_{\mathrm{rms}})^{-1}$.  

\begin{figure*}
\centering
\includegraphics[width=.7\textwidth]{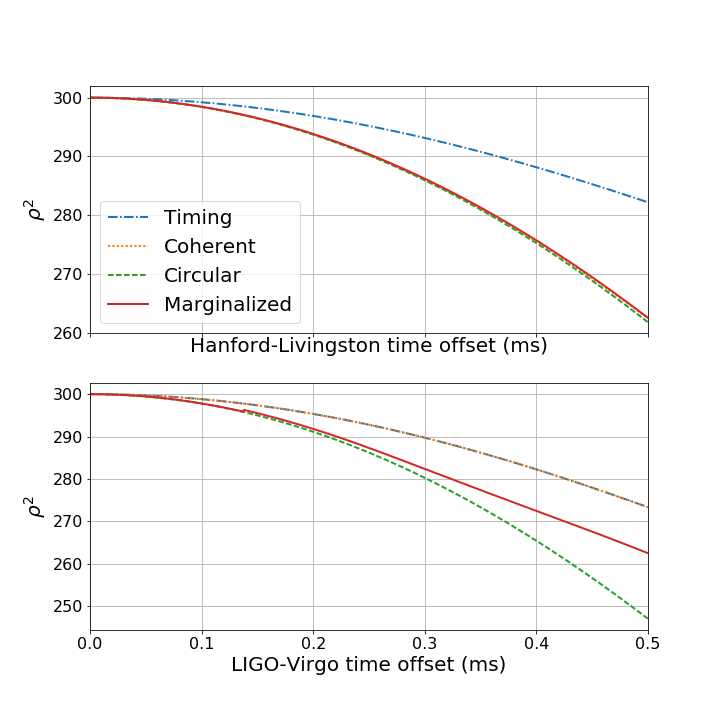}
\caption{The falloff in \ac{SNR} as a function of timing error based on timing,
coherent and circular polarization approximations.  As before, we consider a
source with \ac{SNR} 10 in each of three detectors and introduce a timing offset
between the two LIGO detectors (top panel) or LIGO and Virgo (bottom panel).
We also show the re-weighted \ac{SNR} based on marginalizing over astrophysical
priors.  For a time offset in LIGO, the re-weighted \ac{SNR} matches well with both the coherent and
circular polarization approximations, while for Virgo it matches the circular polarization approximation 
for time offsets up to $0.3$ms.}
\label{fig:like_falloff}
\end{figure*}

To demonstrate the utility of this approximation, \fref{fig:like_falloff} shows the falloff of the 
\ac{SNR} under the timing, coherent and circular polarization approximations and compares these to the 
re-weighted \ac{SNR} obtained from incorporating astrophysical priors and marginalizing over the 
likelihood.  For a relative time delay between the two LIGO detectors, the coherent and circular polarized
results are essentially identical, and match the re-weighted \ac{SNR} result.  Using timing only leads to a 
slower falloff of \ac{SNR} and consequently poorer localization.  For an offset
between LIGO and Virgo, the timing and coherent results are identical, as Virgo
is sensitive to a different polarization.  Restricting to circular polarization causes the \ac{SNR} to fall off 
more quickly, and provides a good match to the re-weighted \ac{SNR}, at least for time offsets
$\lesssim 0.3$ms.  For larger time offsets, the re-weighted \ac{SNR} falls off more slowly,
as at these locations the signal is no longer well approximated as circularly polarized.  We note, however,
that the $90\%$ source localization is determined by points with $\Delta \rho^{2} \le 4.6$ 
and in this region the circular polarization approximation works well.

\section{Localization Results}
\label{sec:results}

\subsection{Advanced LIGO-Virgo network}
\label{sec:cohloc}

A realistic scenario for the evolution of the advanced LIGO and Virgo detectors 
towards their design sensitivities, and the expected localization of sources is given in 
\cite{Aasi:2013wya}.  We will not reproduce the full observing scenario here, but instead present 
results for the advanced LIGO-Virgo network operating at design sensitivity: advanced
LIGO with a sky averaged sensitivity to BNS mergers of 200 Mpc, mean frequency of 100 Hz and 
bandwidth of 120 Hz; Virgo with a sensitivity of 130 Mpc, mean frequency of 130 Hz and 
bandwidth of 150 Hz.  In addition, we assume that all detectors have an 80\% duty cycle and
that the times they operate are independent.

As with previous studies  \cite{Fairhurst2009, Fairhurst:2010is, Singer:2014qca}, we distribute  sources 
uniformly in volume and binary orientation and 
deem sources detectable if the expected network \ac{SNR} is above 12 with an \ac{SNR} above 5 in at 
least two detectors.  Furthermore, we assume 
that a detector will contribute to localization provided the SNR in that detector is greater than 4,
and neglect entirely contributions from detectors with \ac{SNR} less than 4. 
For each source that can be localized by three detectors, we calculate the localization area.
We do this by first calculating the marginalized likelihood, as discussed in \sref{sec:marg_like} and
presented in detail in \ref{sec:marg_details}.  For sources which are essentially face-on, we compute
the localization from \eref{eq:net_prob} by first restricting attention to circularly polarized signals using  \eref{eq:circ_project}.  When the signal is not circularly polarized, we project onto the coherent signal space with 
\eref{eq:projection} when calculating the localization.  
We repeat the calculation for the mirror sky location and combine the distributions from the two
locations as discussed in \sref{sec:mirror}.  The procedure for going from a timing distribution
\eref{eq:net_prob} to an area is presented in \ref{sec:loc_details}.

\begin{figure*}
\centering
\includegraphics[width=.49\textwidth]{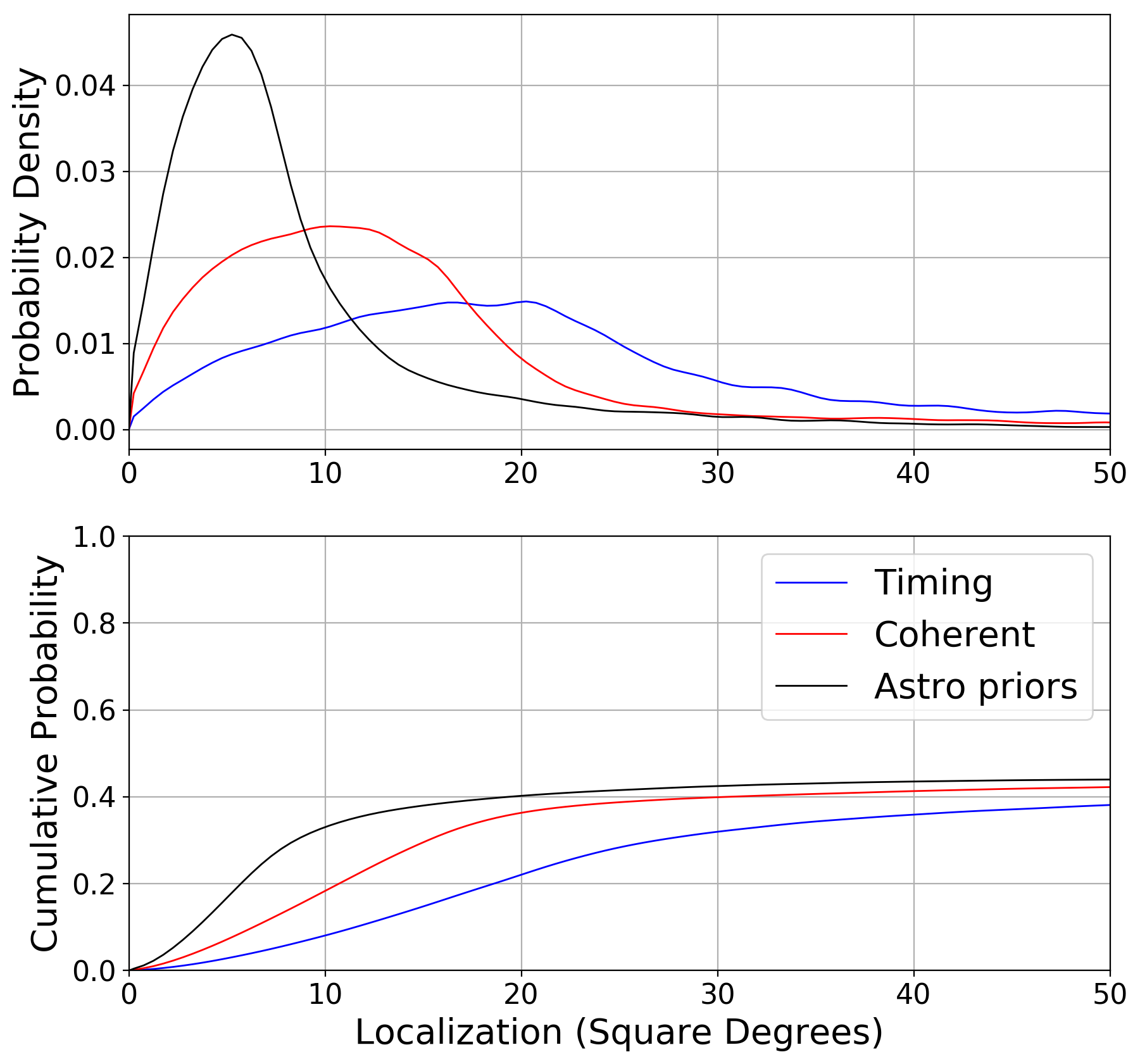} 
\includegraphics[width=.49\textwidth]{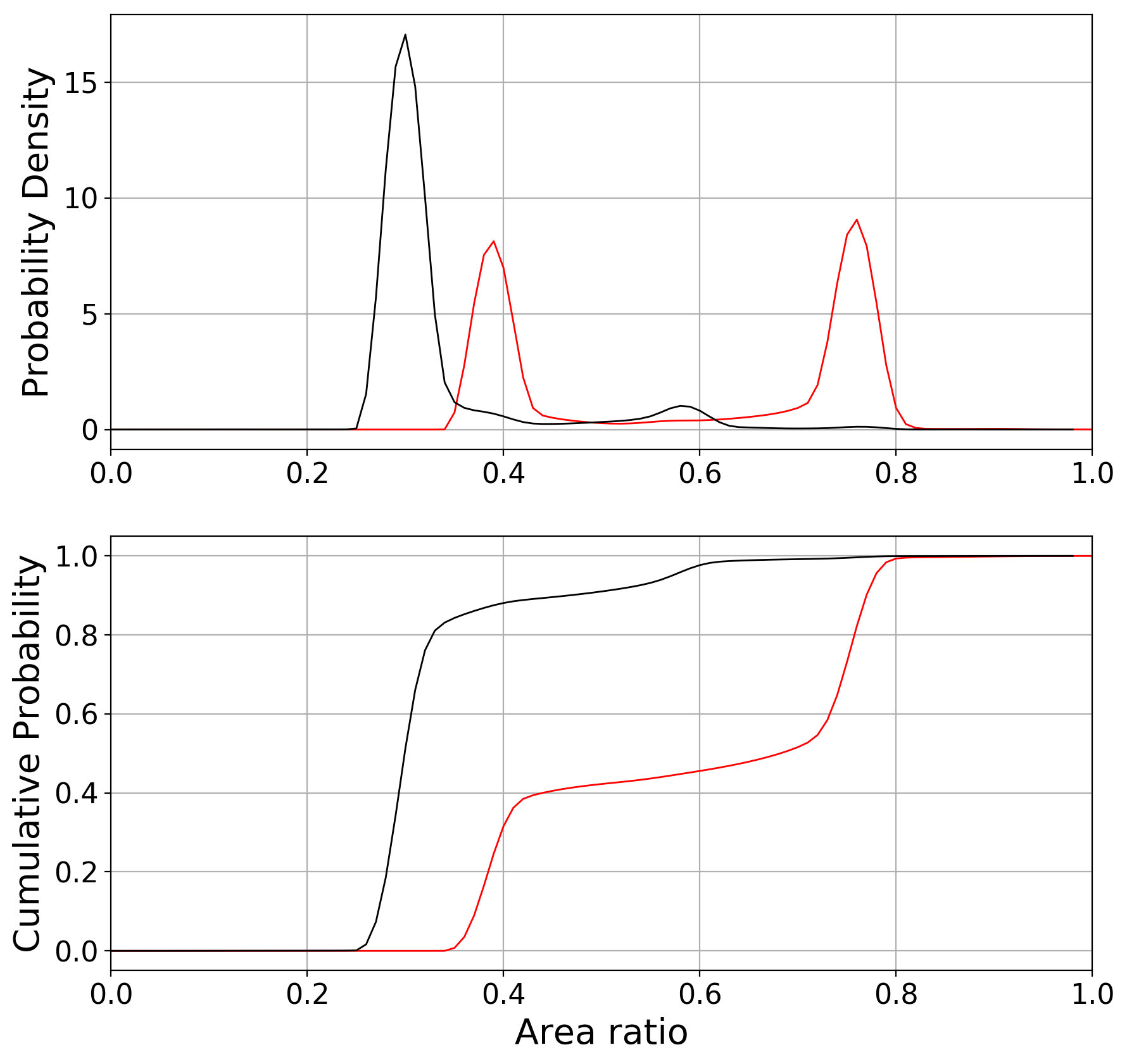}
\caption{Left: The distribution of localization areas for timing and coherent localizations and the use of
astrophysical priors.  The top panel shows the area distribution while the bottom panel gives
the cumulative distribution.  Right: the fractional improvement, on an event by event basis, of
the localization area over the timing result.}
\label{fig:3det_loc}
\end{figure*}

In \fref{fig:3det_loc}, we show the distribution of localization areas for found events, using the
three methods described in \sref{sec:beyond_tri}: timing only, coherent analysis and a coherent analysis
incorporating astrophysically motivated priors.  We show both the distribution of areas and the fractional 
improvements for each event.  The mode of the sky area distribution is reduced from $20 \deg^{2}$ for
timing to $10 \deg^{2}$ for coherent analysis and $6 \deg^{2}$ when we incorporate the expected
astrophysical distribution.  In all cases only around 40\% of sources are localized within $50 \deg^{2}$.  
The majority of events which are not well localized
are seen in only two detectors, either because only 2-detectors are operating due to the assumed duty 
cycle (about 25\%) or the event does not have sufficient \ac{SNR} in one detector for it to contribute to localization 
(about 25\%) with the final 10\% seen with sufficient \ac{SNR} in all detectors to be localized, but just having poor
localization.  

The second panel in \fref{fig:3det_loc} shows the ratio of the coherent or marginalized
area to the value obtained with timing triangulation alone on an event-by-event basis.
This shows two clear peaks for coherent
localization.  The peak at around 0.4 corresponds to events which can be localized to a single sky patch
using coherent localization, while the second peak at around 0.75 includes those events which still give
two sky patches.  In both cases, there is a 25\% reduction in the area of the sky region due to the
requirement of amplitude and phase consistency across the three detectors.
By incorporating astrophysical priors, the localization area for the majority of events is reduced to 30\%
of its original value.  This corresponds to events which are localized in a single sky patch and are effectively
face-on.  The other, smaller peaks can be explained in a similar way: events with 40\% of the original area are 
localized to a single patch, but not face-on; events at 60\% are effectively face-on but from two sky regions;
events at 75\% are not face-on and come from two sky regions.

Finally, we note that in the above plots, we \textit{never} assume that a source can be localized to a
single sky location for the timing results.  This is in contrast to the original work
\cite{Fairhurst2009, Fairhurst:2010is} which gave timing areas under the assumption that sources would
be localized to a single sky patch based on amplitude information.  Thus, the distributions for timing-based 
localization above give twice the area
as those presented in e.g. \cite{Fairhurst2009, Fairhurst:2010is}.
The coherent localization results presented here are comparable to the timing results in the original papers.
However, for each source, the localization is either 25\% better due to 
coherent localization or  50\% worse as the event is localized to two sky patches.  
When we incorporate astrophysical priors, the average localization is 40\% smaller than the single patch 
timing triangulation result.

\subsection{Networks of four and five detectors}
\label{sec:nets}

\begin{figure*}
\centering
\includegraphics[width=.49\textwidth]{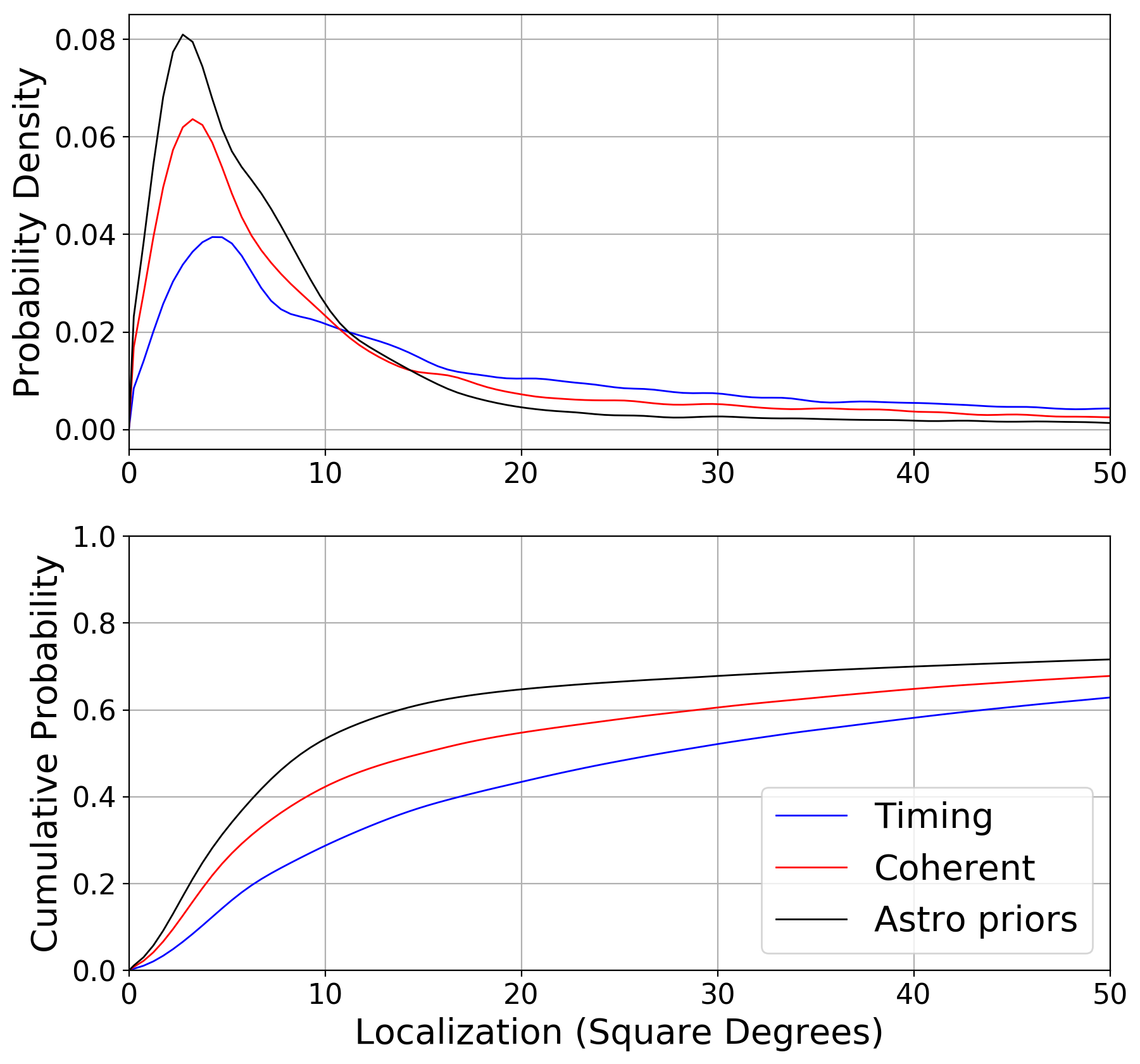} 
\includegraphics[width=.49\textwidth]{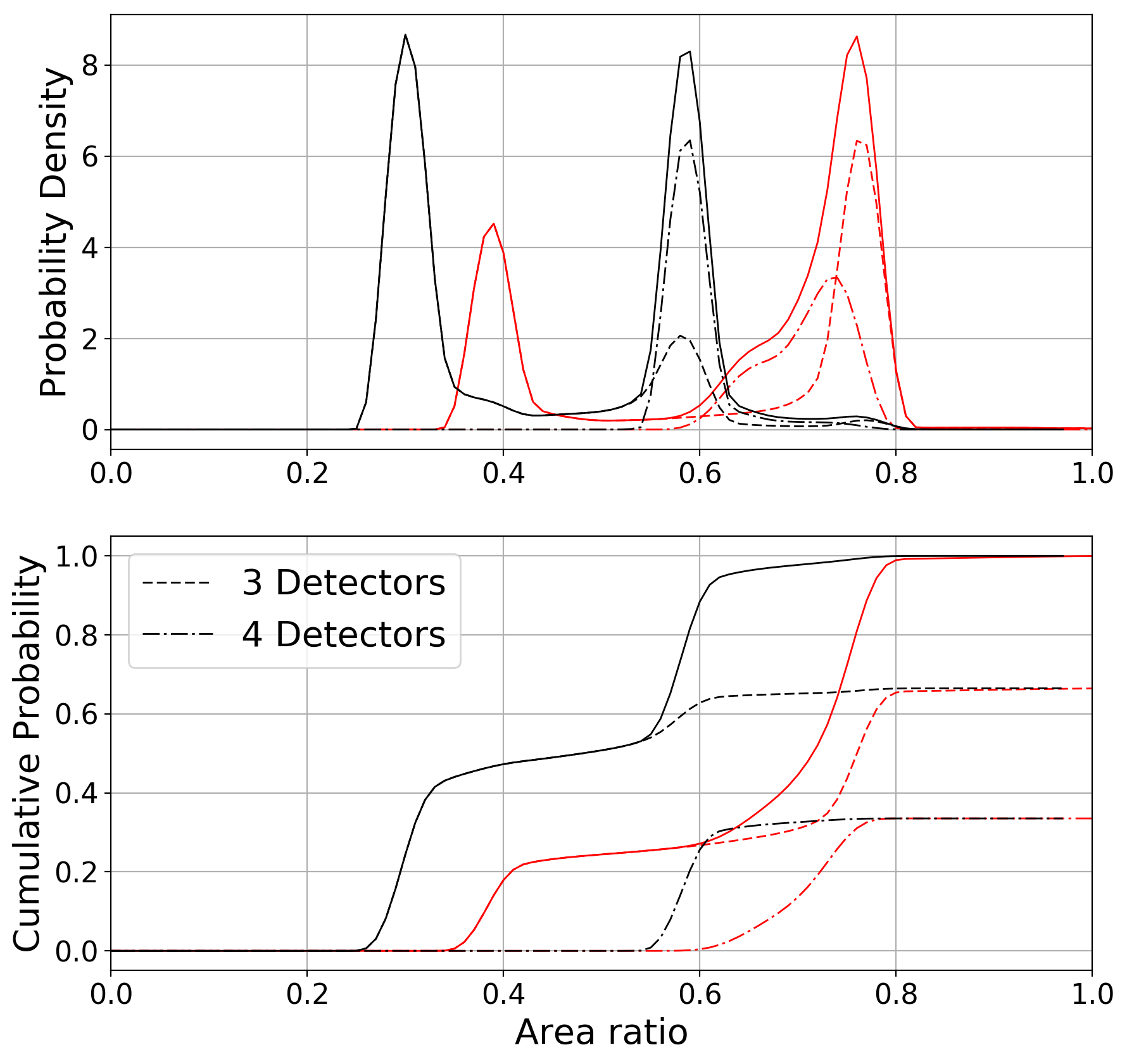}
\caption{Localization of events in the LIGO Hanford--Livingston--India and Virgo network.  
Left: The distribution of localization areas for timing and coherent localizations and the use of
astrophysical priors.  The top panel shows the area distribution while the bottom panel gives
the cumulative distribution.  Right: the fractional improvement, on an event by event basis, of
the localization area over the timing result.  Here, the improvement is sub-divided into
events which are observed in 3 or 4 detectors as there are different factors which affect the
localization in these cases.}
\label{fig:loc_4det}
\end{figure*}

The longer term plan calls for a global network of five detectors, with the addition of KAGRA and LIGO
India \cite{Somiya:2011np, M1100296, Aasi:2013wya}.  The analysis above can easily be extended to 
these networks.  Sky location 
degeneracy is not a problem with four or more detectors --- a unique sky patch can be determined from the 
time delays observed between four detectors.  However, we can still require consistency with two
gravitational wave polarizations, and incorporate astrophysical priors to improve the localization.

In \fref{fig:loc_4det} we show the expected localizations for events observed by the Advanced LIGO-Virgo 
network, where we assume that Advanced LIGO incorporates three detectors, with one in India.  As before,
we show the localization distribution as well as the event-by-event improvement of localization.  The mode 
of the localization distribution decreases from $6 \deg^{2}$ for timing triangulation to $4 \deg^{2}$ for
coherent and $3 \deg^{2}$ when astrophysical priors are incorporated.  Around $70\%$ of found events are 
localized within $50 \deg^{2}$, with the remainder either observed when only two detectors were operational
($7\%$), only seen in two detectors ($11\%$) or observed in three detectors and poorly localized ($12\%$).
Of the events which are well localized, one third are observed in four detectors, due to live time and \ac{SNR}
thresholds, with the remainder observed in three.  Since events observed in four detectors are
localized to a single sky patch based on timing alone, the majority of these events see a $25-35\%$ 
reduction in area from coherent localization and a $40\%$ reduction when we incorporate an appropriate
astrophysical distribution.  The fact that coherent localization has a bigger impact for events observed in four
detectors ($25-35\%$ improvement rather than $25\%$ with three detectors) is expected as this leads to 
a reduction from four data streams, rather than three, to the two \ac{GW} polarizations.

\begin{figure*}
\centering
\includegraphics[width=.49\textwidth]{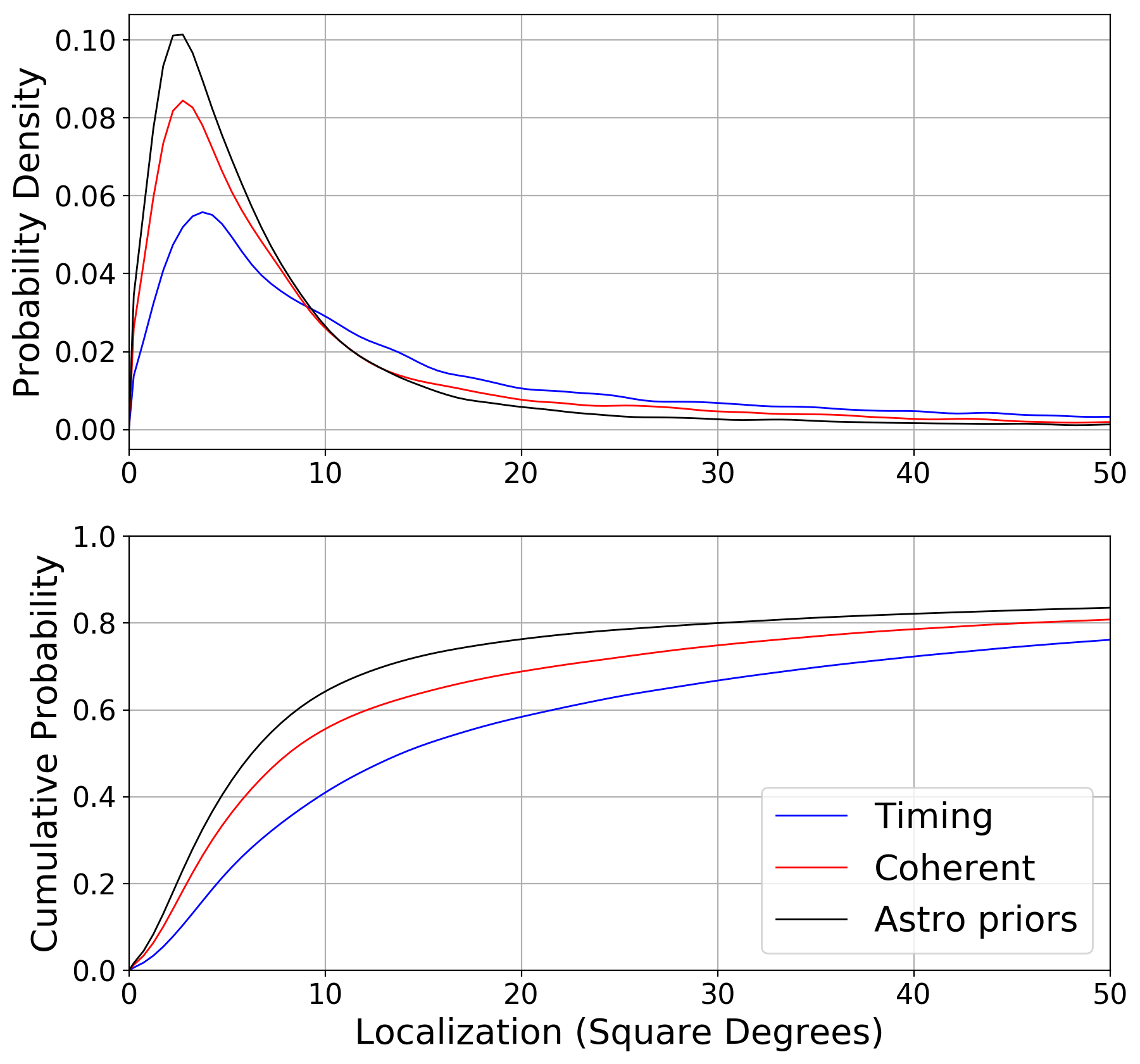} 
\includegraphics[width=.49\textwidth]{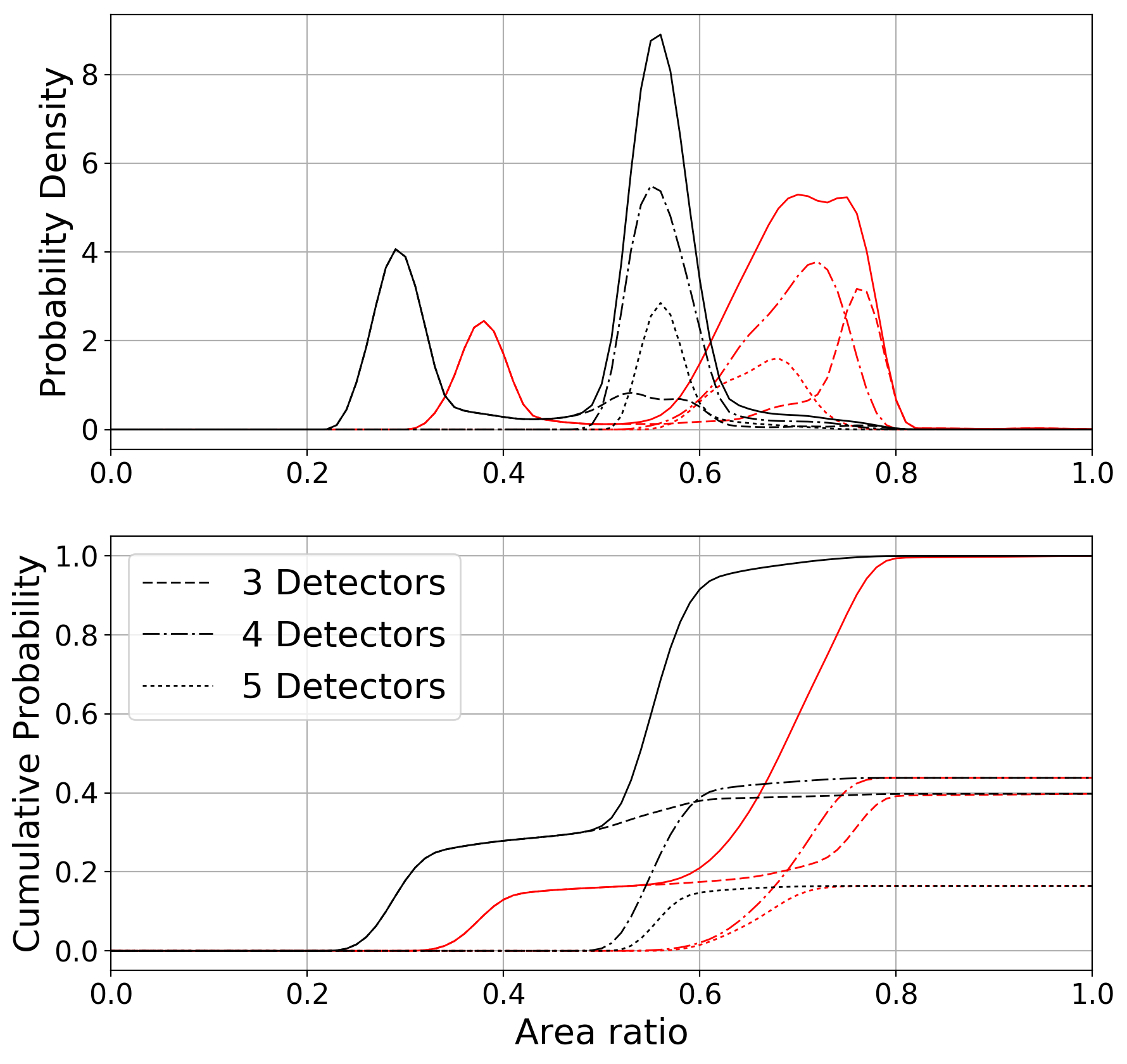}
\caption{Localization of events in the LIGO Hanford--Livingston--India, Virgo, KAGRA network.  
Plots are in \fref{fig:loc_4det} with the fractional improvement in area now divided between 3, 4
and 5 detector observations.}
\label{fig:loc_5det}
\end{figure*}

In \fref{fig:loc_5det} we show the expected localizations for events observed by the 
Advanced LIGO-Virgo-KAGRA network,
where we assume that Advanced LIGO incorporates three detectors, with one in India.  The results
are similar to the four-detector network, with the mode of the area distribution decreasing from 
$4 \deg^{2}$ for timing triangulation to $3 \deg^{2}$ for
coherent and $2 \deg^{2}$ when astrophysical priors are incorporated.  Around 80\% of observed
events are localized within $50\deg^{2}$ with $2\%$ of those not localized observed when only two
detectors are operational; $5\%$ observed in only two detectors and $13\%$ localized more poorly
than $50\deg^{2}$.  
 For the 5-detector network, around 20\% of well-localized
events are seen in all five detectors, with 40\% in four and the remaining 40\% in three.  Qualitatively,
the 5-detector events are similar to those seen in 4-detectors:  they are already localized to a single
sky patch so coherent localization serves to reduce the area by $30-40\%$, where again
the localization improvement is better than with 3 or 4 detectors from consistency with two \ac{GW}
polarizations.  For events seem in four or five detectors, inclusion of astrophysical priors 
reduces the area by between $40\%$ and $50\%$.

\section{Discussion}
\label{sec:disc}

In this paper, we have examined in detail two effects that improve source localization over and above
what is achieved with the timing triangulation approximation.  The first is the requirement of a coherent
signal, comprised of two gravitational wave polarizations, observed consistently in all detectors in the
network.  The second is the incorporation of an astrophysically motivated distribution of events, which
are used as priors in the distance and orientation distributions, and lead to the expectation that the
majority of observed signals will be distant and close to face-on.  Both of these lead to considerable
improvements in localization over timing, in the ability to restrict signals to a single sky patch
and also to reduce the area of the sky patch.  In previous timing triangulation studies, we assumed
that sources could be localized to a single sky patch using additional information.  Here, we have
justified that assumption by showing that the vast majority, over $90\%$, of sources will be localized
to a single sky patch.  Furthermore, the area of the sky region will be around $40\%$ smaller than
predicted using timing information alone.  Finally, we note that the improvement in localization
is not uniform for all events --- it varies depending upon whether events can, indeed, be
localized to a single region and also whether they are recovered as circularly polarized.

In \cite{Grover:2013sha} a detailed comparison between timing triangulation and full parameter
estimation localization was performed.  The results presented here help to explain the discrepancies
observed.  In a future work, we plan to repeat the detailed comparison with full localization
to verify that the new formalism provides a better approximation to the full results, and identify
cases where the approximations introduced here are inadequate to capture the full parameter recovery
results.

In this paper, we have not considered the localization of events observed in only two detectors.  When
only two detectors are operating, timing triangulation can only give localization to a ring on the sky.
The coherent analysis does not improve this --- the observed signal is consistent with a binary
at a given distance and orientation at every sky point in the ring \cite{BosePaiDhurandhar2000}.  
However, for many of these points, the binary will be inferred to be nearby and close to edge-on.  
By incorporating the astrophysical priors, those points are rejected \cite{Raymond:2008im, 
Kasliwal:2013yqa, Singer:2015ema}.  
In addition, even when three detectors are operating, a reasonable
fraction of events will be observed in only two of them.  As demonstrated
by the observation of GW170817 \cite{TheLIGOScientific:2017qsa}, the lack of a clear signal in one 
detector can serve to significantly reduce
the localization uncertainty of the event, as the source can be restricted to originate from a location
close to one of the dead spots for the that detector.  The techniques presented here can be extended
to the case where two detectors observe a signal, and we will investigate localization with two detectors
in a future work.

The analysis presented here is not intended to replace the detailed parameter estimation and
localization techniques  \cite{Singer:2015ema, Veitch:2014wba, Berry:2014jja, 2015arXiv150204370P} 
as it still makes use of a number of approximations.  Nonetheless there are useful applications of an
approximate method.  The calculation of the 
localization is computationally straightforward, which allows for extensive studies to be run
in a straightforward manner.  This may identify regions of parameter space which are interesting
for more detailed follow-up studies.  The method presented here also provides some intuition into the
key factors contributing to localization
and provides a simple sanity check to the results obtained from complex analysis pipelines.
The insights obtained from a simple understanding can also be used to inform development of detailed analyses.
For example, it may be possible to adapt the jump proposals in an MCMC routine to follow the observed
degeneracies in the approximations presented here \cite{Raymond:2014uha}.  
Alternatively, the approximate likelihood 
is a natural candidate for the network detection statistic in an analysis pipeline as it
naturally folds in the astrophysical weighting of the signal.  Indeed, the improved detection statistic
presented in \cite{Nitz:2017svb} was independently motivated by investigations of calculating the
marginalized likelihood for two detector observations.

Another application of this analysis is the study of calibration accuracy requirements for 
current and future detectors.  In \cite{Fairhurst2009} we investigated the impact of calibration
uncertainty on timing-based localization.  It would be relatively straightforward to repeat that analysis
using the improved localization techniques discussed in this paper.  As an example, \fref{fig:inclination} 
makes it clear that
for approximately face-on sources the amplitude of the two gravitational wave polarizations must be nearly equal
and the phase difference essentially $90^{\circ}$.  And, if the source is well localized, the detector response is known
with good accuracy.  Then, we can test whether the observed amplitude
and phase in the detectors in the network are consistent with this expectation.  For example, with 
GW170817, the sky location is known from \ac{EM} observations \cite{GBM:2017lvd} and the \ac{GW} signal
was consistent with circular polarization \cite{Abbott:2017xzu}.  From this information, it is possible to 
calculate the expected relative amplitude and phase differences between the signal observed at the Hanford and 
Livingston detectors.  At the \ac{SNR} that
the event was observed with, this would allow a test of the relative calibration on the order of a few percent
in amplitude and a few degrees in phase.  A related method of using astrophysical observations to
calibrate the detectors has been discussed in \cite{Pitkin:2015kgm}.

A natural application is to the study of localization capabilities of third generation gravitational wave
detector networks.  An initial investigation into the performance of homogeneous and heterogeneous
networks was performed in \cite{Mills:2017urp}.  These networks are expected to be sensitive
to binary neutron star mergers to $z\sim 1$ and binary black hole mergers to $z\sim 20$.  
The merger rate will vary significantly over this range and the approximation of
a uniform volumetric rate no longer holds.  Indeed, as has been pointed out in \cite{Vitale:2016icu}, the 
bias towards face-on sources will not exist for third generation networks.  Consequently, in \cite{Mills:2017urp} we
used coherent localization, without assuming an astrophysical distribution.  However, it should be 
relatively straightforward to incorporate changes in the astrophysical priors due to redshift effects and the 
evolution of the merger rate over the history of the universe to obtain more accurate estimates of source localization
in third generation gravitational wave networks. 
\section*{Acknowledgments}

We would like to thank many people for interesting discussions on this
topic over the years, in particular Duncan Brown, Kipp Cannon, Ilya Mandel, Cameron Mills, 
Larry Price, Frank Ohme, Bangalore Sathyaprakash, Leo Singer, Patrick Sutton, 
Vaibhav Tiwari and John Veitch.  This research was made possible thanks to support from the Royal Society 
and the UK Science and Technology Facilities Council (STFC).

\appendix
\section{Marginalized Likelihood}
\label{sec:marg_details}

In this Appendix, we derive an approximate expression for the likelihood, marginalized over distance and
orientation parameters.  In particular,
we calculate the likelihood from a generic orientation and also from a face-on, or face-off, signal and show
that the combined likelihood is well approximated by the sum of these.
We begin in \ref{sec:signal} by briefly reviewing the properties of the waveform and its parametrization.
In \ref{sec:like}, we derive the marginalized likelihood around the maximum \ac{SNR} and in 
\ref{sec:circ} we obtain the result when restricted to circularly polarized signals (face-on or face-off).
Finally, we give the combined result in \ref{sec:combined}.

\subsection{The binary merger waveform}
\label{sec:signal}

The response of a gravitational wave detector to a signal can be expressed as \cite{thorne.k:1987}
\begin{equation}\label{eq:h_t}
  h(t) = F_{+}(\theta, \phi, \psi) h_{+}(t) + 
         F_{\times}(\theta, \phi, \psi) h_{\times}(t) 
\end{equation}
where $h_{+}$ and $h_{\times}$ are the two polarizations of the
gravitational wave and $F_{+}$ and $F_{\times}$ are the detector response functions that
depend upon location of the source and its polarization.  

The merger waveform can be calculated to high accuracy using the post-Newtonian framework
\cite{lrr-2014-2} during the inspiral phase and numerical relativity simulations when the binary
components are close to merger.  For a generic binary, the waveform can be complicated due to
both the precession of the binary \cite{Apostolatos:1994} and also the contribution of multiple signal 
harmonics to the waveform \cite{Capano:2013raa}.  In this paper, we will not consider either of these 
effects, and restrict our attention to non-precessing systems that have either no spin or spins aligned 
with the binary's orbital angular momentum \cite{Ajith:2007qp} and for which the higher signal harmonics
do not contribute significantly.  This is appropriate for binary neutron star systems, where the spins
of the neutron star do not have a significant effect on the waveform and also for systems of comparable
masses where the spins are close to aligned 
\cite{Brown:BNSSpin, PhysRevD.89.024010, Brown:2012nn, Capano:2013raa}

In this case we can express the waveforms as
\begin{eqnarray}\label{eq:h_plus_cross}
h_{+}(t) &=&  \mathcal{A}^{1} h_{0}(t) + \mathcal{A}^{3}
h_{\frac{\pi}{2}}(t)
\nonumber \\
h_{\times}(t) &=& \mathcal{A}^{2} h_{0}(t) + 
\mathcal{A}^{4} h_{\frac{\pi}{2}}(t) \, ,
\end{eqnarray}
where $h_{0}$ and $h_{\frac{\pi}{2}}$ denote the two phases of the waveform and will depend upon
the masses and (aligned) spins of the binary.  Furthermore, the two phases of the waveform
satisfy
\begin{equation}\label{eq:elliptic}
  \tilde{h}_{\frac{\pi}{2}}(f) = i \tilde{h}_{0}(f) \, .
\end{equation}

The functional form of the 
$\mathcal{A}^{i}$ is well known, and depends upon four free
parameters: the distance $D$, inclination $\iota$, polarization $\psi$ and coalescence
phase $\phi_{0}$ of the system \cite{Cornish:2006ms,BosePaiDhurandhar2000}:

\begin{eqnarray}
    \label{eq:amps}
        \mathcal{A}^1 &= \frac{D_0}{D} \frac{\left(1 + \cos^2 \iota \right)}{2} \cos 2 \phi_0 \cos 2 \psi  
                - \frac{D_0}{D} \cos \iota \sin 2 \phi_0 \sin 2 \psi  \, ,  \nonumber \\
        \mathcal{A}^2 &= \frac{D_0}{D} \frac{\left(1 + \cos^2 \iota \right)}{2} \cos 2 \phi_0 \sin 2 \psi  
                + \frac{D_0}{D} \cos \iota \sin 2 \phi_0 \cos 2 \psi  \, , \nonumber \\
        \mathcal{A}^3 &= -\frac{D_0}{D} \frac{\left(1 + \cos^2 \iota \right)}{2} \sin 2 \phi_0 \cos 2 \psi  
                - \frac{D_0}{D} \cos \iota \cos 2 \phi_0 \sin 2 \psi  \, , \nonumber \\
        \mathcal{A}^4 &= -\frac{D_0}{D} \frac{\left(1 + \cos^2 \iota \right)}{2} \sin 2 \phi_0 \sin 2 \psi 
                + \frac{D_0}{D} \cos \iota \cos 2 \phi_0 \cos 2 \psi  \, .
\end{eqnarray}
Here, $D_{0}$ is a fiducial distance used in generating the waveforms $h_{0}$ and $h_{\frac{\pi}{2}}$.  The
choice of $D_{0}$ is arbitrary and does not affect any of the results presented below.
Furthermore, every set of values for  $\mathcal{A}^{\mu}$ corresponds to a set of physical parameters, 
unique up to an overall reflection symmetry, $\psi \rightarrow \psi + \frac{\pi}{2}$
and $\phi \rightarrow \phi + \frac{\pi}{2}$.

Combining equations (\ref{eq:h_t}) and (\ref{eq:h_plus_cross}), we can express the signal $h$
at a detector $i$ as
\begin{equation}\label{eq:h_i}
h_{i} = (\mathcal{A}^{1} F^{i}_{+} + \mathcal{A}^{2} F^{i}_{\times}) h_{0}(t_{i}) +
(\mathcal{A}^{3} F^{i}_{+} + \mathcal{A}^{4} F^{i}_{\times}) h_{\frac{\pi}{2}}(t_{i})
\end{equation}
where $t_{i}$ is the time of arrival of the signal at the detector and $\mathcal{A}^{\mu}$
are the four amplitudes introduced in \eref{eq:amps}.

\subsection{Marginalized Likelihood}
\label{sec:like}

The probability that a given signal $h$ is present in the data is proportional
to the likelihood.  For a network of detectors, the log-likelihood is simply
the sum of individual detector contributions:
\begin{equation}
	\ln \Lambda = \sum_{i} \ln \Lambda_{i} = \mathbf{(s|h)} - \frac{1}{2} \mathbf{(h|h)}
	\quad \mathrm{where} \quad
	\mathbf{(a|b)} = \sum_{i} (a_{i} |b_{i})
\end{equation}
and the inner product is defined in \eref{eq:inner}.

It has been shown, see e.g.~\cite{Jaranowski:1998qm}, that the likelihood can be maximized over the
four amplitude parameters encoded by the $\mathcal{A}^{\mu}$.
Here, we will re-derive this result as a starting point for marginalizing over the physical
parameters, using a notation that more closely follows that used in 
gravitational wave searches \cite{Babak:2012zx, Usman:2015kfa}.  In particular, we work
with the complex SNR, $Z$, defined in equation (\ref{eq:complex_snr}).  
We combine the $\mathcal{A}^{\mu}$ into two complex parameters
\begin{equation}
	\mathcal{A}^{+} = \mathcal{A}^{1} + i \mathcal{A}^{3} 
	\quad \mathrm{and} \quad
	\mathcal{A}^{\times} = \mathcal{A}^{2} + i \mathcal{A}^{4} \, .
\end{equation}
Then, we can write the inner product between signal, $s$, and template waveform, $h$, as
\begin{equation}
(s | h) = \mathrm{Re}\left[
(\mathcal{A}^{+} w_{+} + \mathcal{A}^{\times} w_{\times})^{\star} Z 
\right] \, ,
\end{equation}
where $w_{+, \times}$ are the sensitivity weighted detector response functions introduced in 
\eref{eq:projection}.

The log-likelihood for the network can then be expressed as
\begin{equation}
\ln \Lambda = \mathrm{Re}\left[
(\mathcal{A}^{\alpha})^{\star}  w^{i}_{\alpha} Z_{i} 
\right] 
- \frac{1}{2} (\mathcal{A}^{\alpha})^{\star} \mathcal{M}_{\alpha \beta} \mathcal{A}^{\beta}
\end{equation}
where there is an implicit sum over both $\alpha \in (+, \times)$  and $i$, which runs over the
detectors in the network. The matrix $\mathcal{M}$ encapsulates the network's sensitivity to 
the two \ac{GW} polarizations, and is defined as
\begin{equation}
\mathcal{M}_{\alpha \beta} = \sum_{i} w^{i}_{\alpha} w^{i}_{\beta} \, .
\end{equation}
It is straightforward to differentiate the likelihood and show that the maximum occurs at
\begin{equation}
\hat{\mathcal{A}}^{\beta} = \mathcal{M}^{\alpha \beta} w^{i}_{\alpha} Z_{i} \, ,
\end{equation}
where $\mathcal{M}^{\alpha \beta}$ is taken to be the inverse of $\mathcal{M}_{\alpha \beta}$
\cite{Prix:2009tq}.  The maximum likelihood is
\begin{equation}\label{eq:like_max}
\ln \Lambda_{\mathrm{max}} = 
\frac{1}{2} (\hat{\mathcal{A}}^{\alpha})^{\star} \mathcal{M}_{\alpha \beta}
\hat{\mathcal{A}}^{\beta} 
= \frac{1}{2} Z_{i}^{\star} P^{ij} Z_{j}
\end{equation}
where $P_{ij}$ is a projection onto the two-dimensional signal space
\begin{equation}
P^{ij} = w^{i}_{\alpha} \mathcal{M}^{\alpha \beta} w^{j}_{\beta} \, .
\end{equation}
In the dominant polarization, where $w_{+} \cdot w_{\times} = 0$, 
$\mathcal{M}_{\alpha \beta}^{\mathrm{DP}} =
\mathrm{diag}(w_{+}^{2}, w_{\times}^{2}, w_{+}^{2}, w_{\times}^{2})$ 
and the projection operator can be expressed as
\begin{equation}
P^{ij}_{\mathrm{DP}} =  \frac{ w^{i}_{+} w^{j}_{+} }{ | w_{+}|^{2} } + 
 \frac{w^{i}_{\times} w^{j}_{\times} }{ | w_{\times}|^{2} } \, .
\end{equation}

The likelihood can be written as
\begin{equation}\label{eq:like_quad}
\ln \Lambda =
\frac{1}{2} (\hat{\mathcal{A}}^{\alpha})^{\star} \mathcal{M}_{\alpha \beta}
\hat{\mathcal{A}}^{\beta} \label{eq:like} - \frac{1}{2} \mathrm{Re} \left[
(\hat{\mathcal{A}}^{\alpha} - \mathcal{A}^{\alpha})^{\star} \mathcal{M}_{\alpha \beta}
(\hat{\mathcal{A}}^{\beta} - \mathcal{A}^{\beta}) \right] \, .
\end{equation}
The first term is the maximum likelihood while the second term shows the quadratic falloff
around the maximum.  The benefit of using the $\mathcal{A}$ variables is clear, as the likelihood 
is a multivariate Gaussian. 

We now wish to marginalize \eref{eq:like_quad} over the four parameters $D, \iota, \psi, \phi$.
Taking a prior that is uniform in the $\mathcal{A}^{\mu}$ the integral can be easily performed.  
As discussed in detail in \cite{Prix:2009tq}, the flat prior on the $\mathcal{A}^{\mu}$ 
corresponds to a highly non-physical prior on the original parameters. In particular, 
\begin{equation}\label{eq:amplitude_prior}
  d\mathcal{A}^{1} d\mathcal{A}^{2} d\mathcal{A}^{3} d\mathcal{A}^{4} =  
  \frac{D_{0}^4 dD}{2 D^{5}} (1 - \cos^{2} \iota)^{3} d\cos\iota d\psi \, d\phi \, ,
\end{equation}
where, as before, $D_{0}$ is the fiducial distance used in defining $h_{0, \frac{\pi}{2}}$.

Simple astrophysical considerations lead us, instead, to choose a prior that is uniform in volume, with source
orientations uniformly distributed on a two-sphere and a uniform distribution in phase, 
\begin{equation}\label{eq:physical_prior}
  \frac{3 D^{2} dD}{D_{\mathrm{max}}^{3}} \frac{d \cos\iota}{2} \frac{d \psi}{2 \pi} \frac{d \phi}{2 \pi} = 
  \frac{3 D^{7}}{4 \pi^{2} D_{0}^4 D_{\mathrm{max}}^{3}} (1 - \cos^{2} \iota)^{-3} 
  d\mathcal{A}^{1} d\mathcal{A}^{2} d\mathcal{A}^{3} d\mathcal{A}^{4} \, , \nonumber
\end{equation}
where $D_{\mathrm{max}}$ is the maximum distance to which the integral is performed.  The normalization is chosen so that
the prior integrates to unity.  Comparing \eref{eq:amplitude_prior} to \eref{eq:physical_prior}, it is clear that
the flat in amplitudes prior significantly over-weights nearby sources
which are close to edge-on, as discussed in \cite{Prix:2009tq, Whelan:2013xka}.  

The exact solution for the marginalized likelihood has been calculated for number of special cases in
\cite{Whelan:2013xka} and was recently computed in general \cite{Dhurandhar:2017rlr}.  Here, we will not use 
the exact expression, but rather look 
to obtain an approximate expression which is accurate enough
to be applied to the localization problem.
The most straightforward method is to re-weight the likelihood by the value of the prior at the 
maximum likelihood values
$\hat{\mathcal{A}}^{\mu}$ \cite{Whelan:2013xka}.  Under this approximation, it is straightforward to 
evaluate the Gaussian integral as
\begin{equation}\label{eq:like_marg}
  \Lambda_{\mathrm{marg}} \approx \frac{24 \hat{D}^{7}}{D_{0}^4 D_{\mathrm{max}}^{3}}
  \frac{(1 - \cos^{2} \hat{\iota})^{-3} }{w_{+}^{2} \, w_{\times}^{2}} \exp\left[\frac{\rho_{\mathrm{net}}^{2}}{2}\right] \, ,
\end{equation}
where the factor of 8 arises due to the fact that the $\mathcal{A}^{\mu}$ are functions of 
$2\phi$ and $2\psi$, giving a factor of 4, and the existence of a discrete degeneracy 
$\phi \rightarrow \phi + \pi/2$ and $\psi \rightarrow \psi + \pi/2$ gives the final factor of two.  
The Gaussian integral evaluates to $4\pi^{2}/ (w_{+}^{2} w_{\times}^{2})$, as can easily be seen 
in the dominant polarization frame.

\subsection{Circularly Polarized Signal}
\label{sec:circ}

Unfortunately, the simple approximation \eref{eq:like_marg} breaks down when the system is close to 
circularly polarized, i.e. $|\cos \iota| \approx 1$, as discussed in detail in \cite{Whelan:2013xka}.  
The reason is that the marginalized likelihood is dominated by contributions away from 
the peak, and so approximating the integral based on the area around the maximum likelihood no longer
works. For a face-on binary, the signal is circularly polarized and there are only two 
remaining degrees of freedom: the overall amplitude and a single phase $2(\phi \pm \psi)$, with the 
sign depending upon if the system is left or right circular.  Thus, the volume of phase space consistent with a 
circularly polarized binary
is significantly larger than for other systems.  Also, for face-on binaries, the inferred distance is the
largest, again increasing the marginalized likelihood.  

Here, we provide a simple approximation to the likelihood for approximately
circularly polarized signals.  
We restrict to circularly polarized waveforms, as described in \cite{Williamson:2014wma}, by 
requiring
\begin{equation}
	\mathcal{A} := \mathcal{A}_{+} = \pm i \mathcal{A}_{\times}  \, ,
\end{equation}
where the positive corresponds to left-circular polarization and negative to right.
The complex amplitude $\mathcal{A}$ can be written as
\begin{equation}
	\mathcal{A} \approx \frac{D_{0}}{D} \cos \iota e^{-i \phi_{L, R} } 
	\quad \mathrm{where} \quad
	\phi_{L,R} = 2(\phi_{0} \pm \psi) \, ,
\end{equation}
and the likelihood as
\begin{equation}
	\ln \Lambda = 
	\mathrm{Re}\left[ \mathcal{A}^{\star}  (w^{i}_{+} \pm i w^{i}_{\times}) Z_{i}  \right] 
	- \frac{1}{2} |\mathcal{A} |^{2} ( | w_{+} |^{2} + | w_{\times} |^{2} )
\end{equation}
As before, it's straightforward to maximize and obtain the value of $\mathcal{A}$ at the peak, 
\begin{equation}
\hat{\mathcal{A}}= \frac{ (w^{i}_{+} \pm i w^{i}_{\times}) Z_{i} }{|w_{+}|^{2} + |w_{\times} |^{2} } \, ,
\end{equation} 
and the maximum likelihood
\begin{equation}\label{eq:like_max_circ}
	\ln \Lambda_{\mathrm{max}} 
	= \frac{1}{2} ( |w_{+}|^{2} + |w_{\times} |^{2}) \hat{\mathcal{A}}^{\star} \hat{ \mathcal{A}}
	= \frac{1}{2} Z_{i}^{\star} P^{ij} Z_{j}
\end{equation}
where $P_{ij}$ is a projection onto the one-dimensional signal space
\begin{equation}
P^{ij} = \frac{ (w_{+}^{i} \pm i w_{\times}^{i})^{\star} (w_{+}^{j} \pm i w_{\times}^{j})}{ 
| w_{+}|^{2} + |w_{\times} |^{2} } \, .
\end{equation}

The likelihood can be expressed as
\begin{equation}\label{eq:like_circ}
\ln \Lambda =
	\frac{1}{2} \left(|w_{+}|^{2} + |w_{\times} |^{2} \right) 
	\left[ \hat{\mathcal{A}}^{\star} \hat{ \mathcal{A}} - 
	(\hat{\mathcal{A}} - \mathcal{A})^{\star} (\hat{\mathcal{A}} - \mathcal{A}) \right] \, .
\end{equation}
This has the form of a multivariate Gaussian, but here in two, rather than
four, dimensions.  As before, we wish to rewrite the astrophysical prior in terms of variables
appearing in the likelihood.  Here, it is natural to use $|\mathcal{A}|$, $\phi_{L,R}$ and $\cos \iota$.  Then,\footnote{The factor of 8 in the Jacobian of the transformation to $\phi_{L, R}$ is compensated
by the fact that the space $\phi_{L, R} \in [0, 2\pi)$ is covered eight times for $\phi_{0}, \psi \in [0, 2\pi)$ 
(see \cite{Whelan:2013xka} for details).}
\begin{eqnarray}
	\frac{3 D^{2} dD}{D_{\mathrm{max}}^{3}} \frac{d \cos\iota}{2} \frac{d \psi}{2 \pi} \frac{d \phi}{2 \pi} = 
	\frac{3 D_{0}^{3} \cos^{3} \iota }{2 D_{\mathrm{max}}^{3} |\mathcal{A}|^{5}} 
	d\cos\iota |\mathcal{A}| d |\mathcal{A}| 
    \frac{d \phi_{L}}{2 \pi} \frac{d \phi_{R}}{2 \pi} \nonumber \, .
\end{eqnarray}
The likelihood depends upon $|\mathcal{A}|$ and one of $\phi_{L, R}$, but is
independent of the second phase angle and $\cos \iota$.  We begin by integrating out the second phase
angle and $\iota$.  While the likelihood is independent of $\iota$, the approximation of circular polarization
is only appropriate for a limited range of $\iota$.  Consider a generic, not necessarily circularly
polarized, signal then the loss in \ac{SNR} from projecting this onto the circular polarization is
\begin{equation}
	\Delta \rho^{2} = \rho_{\mathrm{net}}^{2} - \rho_{\mathrm{circ}}^{2}
	= \left(\frac{D_{0}}{D} \right)^{2} \frac{w_{+}^{2} w_{\times}^{2}}{w_{+}^{2} + w_{\times}^{2}} 
	\frac{(1 \mp \cos \iota)^{4}}{4} \, .
\end{equation}
The expected value of $\rho^{2}$ due solely to noise contributions is equal to the number of degrees of
freedom.  Thus, we allow for $ \Delta \rho^{2}  \le 2$.  For values smaller than this, the loss in \ac{SNR}
from projection to circular polarization will be indistinguishable from noise.  
We then can solve for the minimum value of $\cos \iota$
\begin{equation}
	|\cos \iota|_{\mathrm{min}} =  1 - 
	\left[\frac{2\sqrt{2}}{\rho} \frac{(w_{+}^{2} + w_{\times}^{2})}{w_{+} w_{\times}} \right]^{1/2} \, ,
\end{equation}
subject to the restriction that $|\cos \iota|_{\mathrm{min}} \ge 0$.  This allows us to approximate the
$\iota$ integral as
\begin{equation}
    \frac{1 - |\cos \iota|_{\mathrm{min}}^{4}}{4} \, . \nonumber
\end{equation}
Finally, we can perform the Gaussian integral, and obtain a factor of $2\pi /
(w_{+}^{2} + w_{\times}^{2})$.  Then, evaluating the prior at the peak, using 
$|\hat{\mathcal{A}}| = \frac{D_{0}}{\hat{D}}$ where $\hat{D}$ 
is the distance associated to a face-on signal, we obtain
\begin{equation}\label{eq:like_circ}
  \Lambda_{\mathrm{marg}} \simeq 
  \frac{3}{8} \frac{\hat{D}^{5}}{D_{0}^2 D_{\mathrm{max}}^{3}} 
  \frac{1 - |\cos \iota|_{\mathrm{min}}^{4}}{(w_{+}^{2} + w_{\times}^{2})}
   \exp\left[\frac{\rho_{\mathrm{circ}}^{2}}{2}\right] \, .
\end{equation}

\subsection{Combined Likelihood}
\label{sec:combined}

We approximate the marginalized likelihood by three contributions: one from around the peak likelihood, 
\eref{eq:like_marg},
and one from each of the left and right circularly polarized restrictions, \eref{eq:like_circ}.  As is clear from \eref{eq:like_marg}, the
peak likelihood contribution diverges as $\cos \iota \rightarrow \pm 1$ as the approximations used break down.
To rectify this, we discard the contribution from the peak any time that 
\begin{equation}\label{eq:face_on_condition}
	\rho^{2}_{\mathrm{net}} -  \rho^{2}_{\mathrm{circ}} \le 2 \, .
\end{equation}
As discussed above, when the difference between the peak \ac{SNR} and the circular projection of the \ac{SNR}
is less than two, there is no significant evidence for power in the second polarization so the circular 
approximation is more appropriate.

In \fref{fig:marg_like} in the main text we show the dependence of the marginalized likelihood, expressed as
a re-weighted \ac{SNR}, as a function of the inclination angle for a source with fixed \ac{SNR}.  When the 
inclination angle is less than $55^{\circ}$ the marginalized likelihood, calculated numerically, is in good
agreement with the circular polarization and the restriction \eref{eq:face_on_condition} is met so that the
peak likelihood doesn't contribute at all.  
For angles up to $40^{\circ}$ the marginalized likelihood
is constant, as the signal is circularly polarized to a very good approximation.  Above $40^{\circ}$ the 
signal is no longer perfectly circular and projecting onto the circular polarization leads to a loss of \ac{SNR},
which is seen from the face-on contribution decreasing rapidly. There is a small 
discontinuity in the likelihood at around $55^{\circ}$ as we transition to including the contribution around the 
maximum likelihood.  For angles above $70^{\circ}$ the likelihood is well approximated by the contribution 
around the maximum likelihood.  For the edge-on system the re-weighted \ac{SNR}
is reduced from 300 to 283.  Since the likelihood, and posterior probability, is proportional to $\exp[\rho^{2}/2]$,
the edge-on signal is down-weighted by a factor of almost 5,000 relative to a face-on signal.   
\section{Localization}
\label{sec:loc_details}

In this appendix, we present the details of the localization calculation.  Furthermore, we show how the 
posterior distribution on time delays can be converted to a localization distribution over the sky.

\subsection{Timing accuracy}
\label{sec:net_time}

To calculate the leading order localisation expression, we expand network \ac{SNR} \eref{eq:coh_snr}
\begin{equation}
\rho_{\mathrm{net}}^{2} = \sum_{i,j} Z_{i}^{\star} P^{ij} Z_{j} \, .
\end{equation}
to quadratic order in $dt_{i}$, using the expression in equation (\ref{eq:z_dt}).  
As discussed in the main text, we neglect changes in the projection operator $P^{ij}$ as we argue
they will be less significant than changes in \ac{SNR}.
When the observed SNRs are consistent with a signal, the projection 
operator is idempotent, i.e.
\begin{equation}\label{eq:proj}
P^{ij} Z_{j} = Z^{i}
\end{equation}
However, in the presence of noise, or when reconstructing the signal from an incorrect sky 
location, this will no longer be the case.  Consequently, we keep the calculation below general,
and require only that the projection operator be Hermitian.

Inserting the expansion of  for $Z(dt)$ in powers of $dt$ \eref{eq:z_dt} into \eref{eq:coh_snr} 
and keeping terms up to second order in $dt$, we obtain
\begin{eqnarray}\label{eq:coh_loc}
\rho_{\mathrm{net}}^{2} (\mathbf{dt}) &=& \sum_{i, j} Z_{i}^{\star} P^{ij} Z_{j}
  + 4\pi \, \mathrm{Im} \left[ \sum_{i,j} (\bar{f_{i}} dt_{i}) Z^{\star}_{i} P^{ij} Z_{j} \right] \nonumber \\
  && - 4\pi^{2} \, \mathrm{Re} \left[ \sum_{i,j} Z^{\star}_{i} P^{ij} Z_{j} (\overline{f^{2}_{i}} dt^{2}_{i} 
  - \bar{f_{i}} dt_{i} \bar{f_{j}} dt_{j}) \right] \nonumber \\
  &=:& \rho_{\mathrm{net}}^{2} + B^{i} dt_{i} - C^{ij} dt_{i} dt_{j} \, ,
\end{eqnarray}
where we have used the fact that the projection operator is Hermitian to simplify the result.  In the above, we have
defined
\begin{eqnarray}
  \rho_{\mathrm{net}}^{2} &=& \sum_{i, j} Z_{i}^{\star} P^{ij} Z_{j} \\
  B^{i} &=& 4\pi \, \mathrm{Im} \left[ \sum_{j} \bar{f_{i}} Z^{\star}_{i} P^{ij} Z_{j} \right]  \label{eq:loc_a}\\
  C^{ij} & = & 4\pi^{2} \, \mathrm{Re} \left[ \left(\sum_{k} Z^{\star}_{i} P^{ik} Z_{k} \right) \overline{f^{2}_{i}} \delta_{ij}
  - (Z^{\star}_{i} P^{ij} Z_{j}) \bar{f_{i}} \bar{f_{j}} \right] \label{eq:loc_c}
\end{eqnarray}

The three quantities in (\ref{eq:coh_loc}) have simple interpretations.  The first, $\rho_{\mathrm{net}}^2$ is the 
network SNR for a source at the original position (i.e. with $dt_{i} = 0$).  If the observed SNRs are
consistent with a signal, i.e. eq.~(\ref{eq:proj}) holds, then $B^{i}$ vanishes as we are taking the imaginary 
part of a real quantity.  If the single detector \acp{SNR} are not consistent with a source, then the maximum 
network \ac{SNR} can be offset from $dt_{i} = 0$: a nearby sky location may give a larger coherent SNR 
\textit{even though the single detector SNRs will be reduced}.  The third term, $C_{ij}$ determines the
 quadratic falloff in \ac{SNR} away from the peak.
 
Restricting to the case where the observed \acp{SNR} $Z_{i}$ are consistent with a signal from the given
sky location, \eref{eq:coh_loc} reduces to the expression \eref{eq:net_prob} given in \sref{sec:coherent}.

\subsection{Sky Localization}
\label{sec:loc}

To convert the timing expression, \eref{eq:coh_loc} to localization area, we
use the same approach as described in \cite{Fairhurst2009, Fairhurst:2010is}.  In this paper,
we simulate signals from a given sky location $\mathbf{r}$ and at a given geocentric
arrival time $t_{0}$.  From these, we can construct the arrival time in each detector as\begin{equation}
	t_{i} = t_{o} - \mathbf{r} \cdot \mathbf{D}_{i} \, ,
\end{equation}
where $\mathbf{D}_{i}$ is the location of the ith detector.\footnote{As discussed in the Appendix of \cite{Fairhurst:2010is}, given an observed set of \acp{SNR}
and times of arrival $t_{i}$, we can use a constrained maximization to calculate the best fit 
$\mathbf{r}$ and $t_{0}$ and from these the best fit arrival times $t_{i}$ consistent
with a physical source.  We then use this location to calculate the projection operator in the following.}
Furthermore, we use the known
location to calculate the projection operator $P^{ij}$ and, in the case of three detector
networks, the projection at the mirror location $P_{\mathrm{mirror}}^{ij}$.  We then calculate 
the quantities $B_{i}$ and $C_{ij}$  from \eref{eq:loc_a} and \eref{eq:loc_c} 
which give the distribution for the arrival times $T_{i}$ as
\begin{equation}\label{eq:time_prob}
	p(T_{i} | t_{i}, Z_{i}) \propto p(T_{i}) \exp \left[ \frac{1}{2} \left\{B^{i} (T_{i} - t_{i}) 
	- C^{ij} (T_{i} - t_{i})(T_{j} - t_{j}) \right\} \right]\, .
\end{equation}
We are interested in re-expressing this in terms of a distribution for the the sky location, $\mathbf{R}$
and geocentric arrival time $T_{0}$, which are related to the arrival times by 
\begin{equation}
	T_{i} = T_{o} - \mathbf{R} \cdot \mathbf{D}_{i} \, .
\end{equation}
For a simulated signal, the observed \acp{SNR} will be consistent with a signal from the sky location
$\mathbf{r}$.  However, they will not necessarily be consistent with either a circularly polarized
signal from that sky location, which we will use in approximating localizations, or with a signal
from the mirror location $\mathbf{r}_{\mathrm{mirror}}$.  In this case, the centre of the 
reconstructed localization region will be offset from the true (or mirror) location and the
peak \ac{SNR} will be lower than the total \ac{SNR} observed in the network.  
In particular, we obtain a maximum coherent \ac{SNR} with time offsets 
\begin{equation}\label{eq:dt_hat}
	 (\hat{t}_{i} - t_{i}) = \frac{1}{2} C^{-1}_{ij} B^{j} \, ,
\end{equation}
which gives the maximum \ac{SNR} as
\begin{equation}\label{eq:coh_max}
	\hat{\rho}_{\mathrm{coh}}^{2} := \rho_{coh}^2 + C^{ij} (\hat{t}_{i} - t_{i}) (\hat{t}_{i} - t_{i})
\end{equation}
Using the time offsets that maximize the \ac{SNR}, we construct a best fit sky location and arrival time
\begin{equation}
	\hat{t}_{i} = \hat{t}_{0} - \hat{\mathbf{r}} \cdot \mathbf{D}_{i}
\end{equation}
which allows us to re-express the localization distribution as
\begin{equation}
	p(T_{0}, \mathbf{R} | t_{i}, Z_{i}) \propto p(T_{0}, \mathbf{R}) \exp \left[ - \frac{1}{2}  
	C^{ij} dt_{i} dt_{j}  \right]\, .
\end{equation}
where
\begin{equation}
	dt_{i} = T_{0} - \hat{t}_{0} - (\mathbf{R} - \hat{\mathbf{r}}) \cdot \mathbf{D}_{i}
\end{equation}

Finally, we would like to eliminate $T_{o}$ to obtain the distribution over the sky.  Since the 
time of arrival will not be known, we simply marginalize over it using a flat prior on $T_{0}$.
In this case, marginalizing the distribution over $T_{o}$ is equivalent
to maximizing the SNR over this variable (details in the appendix of \cite{Fairhurst:2010is}).
Upon doing so, we find that the localization distribution can be expressed as: 
\begin{equation}\label{eq:loc_prob}
	p(T_{0}, \mathbf{R} | t_{i}, Z_{i}) \propto p(\mathbf{R}) \exp \left[ 
	- \frac{1}{2}  
	(\hat{\mathbf{r}} - \mathbf{R})^{T} \mathbf{M} (\hat{\mathbf{r}} - \mathbf{R})
	\right]
\end{equation}
where the matrix $\mathbf{M}$ is given as
\begin{equation}\label{eq:coh_m}
  \mathbf{M} = \frac{1}{2} \sum_{i, j} \mathbf{D_{ij}} \mathbf{D_{ij}}^{T} 
  \left[\frac{c^{i} c^{j}}{c} - C^{ij} \right]
\end{equation}
and
\begin{eqnarray}
c^{i} &=& \sum_{j} C^{ij} 
\quad \mathrm{and} \quad c = \sum_{i} c^{i} \, .
\end{eqnarray}

The localization matrix is very similar in form to the one obtained for timing only.  The main difference
is that there are terms involving correlations between the detectors which were not present before.
It is then reassuring to note that the triangulation expression can be recovered by the replacement 
$P_{ij} \rightarrow \delta_{ij}$.  In this case $A_{i}$ vanishes and 
$C_{ij} = \delta_{ij} c_{i}$, and $c^{i} = (2\pi \rho_{i} \sigma_{f,i})^{2}$.

\section*{References} 

\bibliographystyle{iopart-num}
\bibliography{cbc-group,refs}

\providecommand{\newblock}{}
\begin{thebibliography}{10}
\expandafter\ifx\csname url\endcsname\relax
  \def\url#1{{\tt #1}}\fi
\expandafter\ifx\csname urlprefix\endcsname\relax\def\urlprefix{URL }\fi
\providecommand{\eprint}[2][]{\url{#2}}

\bibitem{Abbott:2016blz}
Abbott B~P {\em et~al.\/} (Virgo, LIGO Scientific) 2016 {\em Phys. Rev.
  Lett.\/} {\bf 116} 061102 (\textit{Preprint} \eprint{1602.03837})

\bibitem{Abbott:2016nmj}
Abbott B~P {\em et~al.\/} (Virgo, LIGO Scientific) 2016 {\em Phys. Rev.
  Lett.\/} {\bf 116} 241103 (\textit{Preprint} \eprint{1606.04855})

\bibitem{TheLIGOScientific:2016pea}
Abbott B~P {\em et~al.\/} (Virgo, LIGO Scientific) 2016 {\em Phys. Rev.\/} {\bf
  X6} 041015 (\textit{Preprint} \eprint{1606.04856})

\bibitem{TheLIGOScientific:2017qsa}
Abbott B~P {\em et~al.\/} (Virgo, LIGO Scientific) 2017 {\em Phys. Rev.
  Lett.\/} {\bf 119} 161101 (\textit{Preprint} \eprint{1710.05832})

\bibitem{Messick:2016aqy}
Messick C {\em et~al.\/} 2017 {\em Phys. Rev.\/} {\bf D95} 042001
  (\textit{Preprint} \eprint{1604.04324})

\bibitem{Usman:2015kfa}
Usman S~A {\em et~al.\/} 2016 {\em Class. Quant. Grav.\/} {\bf 33} 215004
  (\textit{Preprint} \eprint{1508.02357})

\bibitem{Nitz:2017svb}
Nitz A~H, Dent T, Dal~Canton T, Fairhurst S and Brown D~A 2017 {\em Astrophys.
  J.\/} {\bf 849} 118 (\textit{Preprint} \eprint{1705.01513})

\bibitem{Singer:2015ema}
Singer L~P and Price L~R 2016 {\em Phys. Rev.\/} {\bf D93} 024013
  (\textit{Preprint} \eprint{1508.03634})

\bibitem{Veitch:2014wba}
Veitch J, Raymond V, Farr B, Farr W, Graff P, Vitale S, Aylott B, Blackburn K,
  Christensen N, Coughlin M, Del~Pozzo W, Feroz F, Gair J, Haster C~J, Kalogera
  V, Littenberg T, Mandel I, O'Shaughnessy R, Pitkin M, Rodriguez C, R\"over C,
  Sidery T, Smith R, Van Der~Sluys M, Vecchio A, Vousden W and Wade L 2015 {\em
  Phys. Rev. D\/} {\bf 91}(4) 042003 (\textit{Preprint} \eprint{1409.7215})

\bibitem{Monitor:2017mdv}
Abbott B~P {\em et~al.\/} (Virgo, Fermi-GBM, INTEGRAL, LIGO Scientific) 2017
  {\em Astrophys. J.\/} {\bf 848} L13 (\textit{Preprint} \eprint{1710.05834})

\bibitem{GBM:2017lvd}
Abbott B~P {\em et~al.\/} (GROND, SALT Group, OzGrav, DFN, INTEGRAL, Virgo,
  Insight-Hxmt, MAXI Team, Fermi-LAT, J-GEM, RATIR, IceCube, CAASTRO, LWA,
  ePESSTO, GRAWITA, RIMAS, SKA South Africa/MeerKAT, H.E.S.S., 1M2H Team,
  IKI-GW Follow-up, Fermi GBM, Pi of Sky, DWF (Deeper Wider Faster Program),
  Dark Energy Survey, MASTER, AstroSat Cadmium Zinc Telluride Imager Team,
  Swift, Pierre Auger, ASKAP, VINROUGE, JAGWAR, Chandra Team at McGill
  University, TTU-NRAO, GROWTH, AGILE Team, MWA, ATCA, AST3, TOROS, Pan-STARRS,
  NuSTAR, ATLAS Telescopes, BOOTES, CaltechNRAO, LIGO Scientific, High Time
  Resolution Universe Survey, Nordic Optical Telescope, Las Cumbres Observatory
  Group, TZAC Consortium, LOFAR, IPN, DLT40, Texas Tech University, HAWC,
  ANTARES, KU, Dark Energy Camera GW-EM, CALET, Euro VLBI Team, ALMA) 2017 {\em
  Astrophys. J.\/} {\bf 848} L12 (\textit{Preprint} \eprint{1710.05833})

\bibitem{Fairhurst2009}
Fairhurst S 2009 {\em New Journal of Physics\/} {\bf 11} 123006

\bibitem{Fairhurst:2010is}
Fairhurst S 2011 {\em Class. Quantum Grav.\/} {\bf 28} 105021
  (\textit{Preprint} \eprint{1010.6192})

\bibitem{Singer:2014qca}
Singer L~P, Price L~R, Farr B, Urban A~L, Pankow C {\em et~al.\/} 2014 {\em
  Astrophys.J.\/} {\bf 795} 105 (\textit{Preprint} \eprint{1404.5623})

\bibitem{Berry:2014jja}
Berry C~P~L {\em et~al.\/} 2015 {\em Astrophys. J.\/} {\bf 804} 114
  (\textit{Preprint} \eprint{1411.6934})

\bibitem{Grover:2013sha}
Grover K, Fairhurst S, Farr B~F, Mandel I, Rodriguez C {\em et~al.\/} 2014 {\em
  Phys.Rev.\/} {\bf D89} 042004 (\textit{Preprint} \eprint{1310.7454})

\bibitem{Wen:2010cr}
Wen L and Chen Y 2010 {\em Phys.Rev.\/} {\bf D81} 082001 (\textit{Preprint}
  \eprint{1003.2504})

\bibitem{Wen:2005ui}
Wen L and Schutz B~F 2005 {\em Class.Quant.Grav.\/} {\bf 22} S1321--S1336
  (\textit{Preprint} \eprint{gr-qc/0508042})

\bibitem{M1100296}
{Iyer} B {\em et~al.\/} 2011 {LIGO India} Tech. Rep. LIGO-M1100296
  https://dcc.ligo.org/LIGO-M1100296/public

\bibitem{2013IJMPD..2241010U}
{Unnikrishnan} C~S 2013 {\em International Journal of Modern Physics D\/} {\bf
  22} 1341010

\bibitem{Somiya:2011np}
Somiya K (KAGRA) 2012 {\em Class. Quant. Grav.\/} {\bf 29} 124007
  (\textit{Preprint} \eprint{1111.7185})

\bibitem{Harry:2010fr}
Harry I~W and Fairhurst S 2011 {\em Phys. Rev.\/} {\bf D83} 084002
  (\textit{Preprint} \eprint{1012.4939})

\bibitem{Klimenko:2011}
Klimenko S, Vedovato G, Drago M, Mazzolo G, Mitselmakher G, Pankow C, Prodi G,
  Re V, Salemi F and Yakushin I 2011 {\em Phys. Rev. D\/} {\bf 83}(10) 102001
  \urlprefix\url{http://link.aps.org/doi/10.1103/PhysRevD.83.102001}

\bibitem{thorne.k:1987}
Thorne K~S 1987 {\em Three hundred years of gravitation\/} ed Hawking S~W and
  Israel W (Cambridge: Cambridge University Press) chap~9, pp 330--458

\bibitem{GuerselTinto1989}
Guersel Y and Tinto M 1989 {\em \prd\/} {\bf 40} 3884--3938

\bibitem{Keppel:2013uma}
Keppel D 2013  (\textit{Preprint} \eprint{1307.4158})

\bibitem{LIGOS3S4Galaxies}
Kopparapu R~K, Hanna C, Kalogera V, O'Shaughnessy R, Gonzalez G, Brady P~R and
  Fairhurst S 2008 {\em \apj\/} {\bf 675} 1459--1467

\bibitem{TheLIGOScientific:2016htt}
Abbott B~P {\em et~al.\/} (Virgo, LIGO Scientific) 2016 {\em Astrophys. J.\/}
  {\bf 818} L22 (\textit{Preprint} \eprint{1602.03846})

\bibitem{Prix:2009tq}
Prix R and Krishnan B 2009 {\em Class. Quant. Grav.\/} {\bf 26} 204013
  (\textit{Preprint} \eprint{0907.2569})

\bibitem{Whelan:2013xka}
Whelan J~T, Prix R, Cutler C~J and Willis J~L 2014 {\em Class. Quant. Grav.\/}
  {\bf 31} 065002 (\textit{Preprint} \eprint{1311.0065})

\bibitem{Aasi:2013wya}
Abbott B~P {\em et~al.\/} (LIGO Scientific Collaboration, Virgo Collaboration)
  2016 {\em Living Rev. Relat.\/} {\bf 19} 1 (\textit{Preprint}
  \eprint{1304.0670})

\bibitem{BosePaiDhurandhar2000}
Bose S, Pai A and Dhurandhar S~V 2000 {\em Int. J. Mod. Phys.\/} {\bf D9}
  325--329 (\textit{Preprint} \eprint{gr-qc/0002010})

\bibitem{Raymond:2008im}
Raymond V, van~der Sluys M~V, Mandel I, Kalogera V, Rover C and Christensen N
  2009 {\em Class. Quant. Grav.\/} {\bf 26} 114007 (\textit{Preprint}
  \eprint{0812.4302})

\bibitem{Kasliwal:2013yqa}
Kasliwal M~M and Nissanke S 2014 {\em Astrophys. J.\/} {\bf 789} L5
  (\textit{Preprint} \eprint{1309.1554})

\bibitem{2015arXiv150204370P}
{Pankow} C, {Brady} P, {Ochsner} E and {O'Shaughnessy} R 2015 {\em ArXiv
  e-prints\/} (\textit{Preprint} \eprint{1502.04370})

\bibitem{Raymond:2014uha}
Raymond V and Farr W 2014  (\textit{Preprint} \eprint{1402.0053})

\bibitem{Abbott:2017xzu}
Abbott B~P {\em et~al.\/} (LIGO Scientific, VINROUGE, Las Cumbres Observatory,
  DLT40, Virgo, 1M2H, MASTER) 2017 {\em Nature\/} (\textit{Preprint}
  \eprint{1710.05835})

\bibitem{Pitkin:2015kgm}
Pitkin M, Messenger C and Wright L 2016 {\em Phys. Rev.\/} {\bf D93} 062002
  (\textit{Preprint} \eprint{1511.02758})

\bibitem{Mills:2017urp}
Mills J, Tiwari V and Fairhurst S 2017  (\textit{Preprint} \eprint{1708.00806})

\bibitem{Vitale:2016icu}
Vitale S and Evans M 2017 {\em Phys. Rev.\/} {\bf D95} 064052
  (\textit{Preprint} \eprint{1610.06917})

\bibitem{lrr-2014-2}
Blanchet L 2014 {\em Living Reviews in Relativity\/} {\bf 17}
  \urlprefix\url{http://www.livingreviews.org/lrr-2014-2}

\bibitem{Apostolatos:1994}
Apostolatos T~A, Cutler C, Sussman G~J and Thorne K~S 1994 {\em \prd\/} {\bf
  49} 6274

\bibitem{Capano:2013raa}
Capano C, Pan Y and Buonanno A 2014 {\em Phys. Rev.\/} {\bf D89} 102003
  (\textit{Preprint} \eprint{1311.1286})

\bibitem{Ajith:2007qp}
Ajith P {\em et~al.\/} 2007 {\em Class. Quantum Grav.\/} {\bf 24} S689--S699
  (\textit{Preprint} \eprint{arXiv:0704.3764})

\bibitem{Brown:BNSSpin}
{Brown} D~A, {Harry} I, {Lundgren} A and {Nitz} A~H 2012 {\em \prd\/} {\bf 86}
  084017 (\textit{Preprint} \eprint{1207.6406})

\bibitem{PhysRevD.89.024010}
Harry I~W, Nitz A~H, Brown D~A, Lundgren A~P, Ochsner E and Keppel D 2014 {\em
  Phys. Rev. D\/} {\bf 89}(2) 024010
  \urlprefix\url{http://link.aps.org/doi/10.1103/PhysRevD.89.024010}

\bibitem{Brown:2012nn}
Brown D~A, Kumar P and Nitz A~H 2013 {\em Phys.Rev.\/} {\bf D87} 082004
  (\textit{Preprint} \eprint{1211.6184})

\bibitem{Cornish:2006ms}
Cornish N~J and Porter E~K 2007 {\em Class. Quant. Grav.\/} {\bf 24} 5729--5755
  (\textit{Preprint} \eprint{gr-qc/0612091})

\bibitem{Jaranowski:1998qm}
Jaranowski P, Krolak A and Schutz B~F 1998 {\em Phys.Rev.\/} {\bf D58} 063001
  (\textit{Preprint} \eprint{gr-qc/9804014})

\bibitem{Babak:2012zx}
Babak S, Biswas R, Brady P, Brown D, Cannon K {\em et~al.\/} 2013 {\em Phys.
  Rev. D\/} {\bf 87} 024033 (\textit{Preprint} \eprint{1208.3491})

\bibitem{Dhurandhar:2017rlr}
Dhurandhar S, Krishnan B and Willis J~L 2017  (\textit{Preprint}
  \eprint{1707.08163})

\bibitem{Williamson:2014wma}
Williamson A~R, Biwer C, Fairhurst S, Harry I~W, Macdonald E, Macleod D and
  Predoi V 2014 {\em Phys. Rev.\/} {\bf D90} 122004 (\textit{Preprint}
  \eprint{1410.6042})

\end{thebibliography}

\end{document}